\documentclass[a4paper,fleqn]{cas-sc}

\usepackage[authoryear]{natbib}
\usepackage{siunitx}
\usepackage{placeins}
\usepackage{printlen}
\usepackage{nomencl}
\usepackage{setspace}
\makenomenclature
\usepackage{listings}
\usepackage{xcolor}

\definecolor{terminalbg}{RGB}{250,250,250}   
\definecolor{termblue}{RGB}{0,92,197}        
\definecolor{termgreen}{RGB}{34,139,34}      
\definecolor{termgray}{RGB}{120,120,120}     
\definecolor{termorange}{RGB}{200,80,0}      

\lstdefinestyle{terminal}{
  language=Python,
  backgroundcolor=\color{terminalbg},
  basicstyle=\ttfamily\small,
  keywordstyle=\color{termblue}\bfseries,
  stringstyle=\color{termgreen},
  commentstyle=\color{termgray}\itshape,
  numberstyle=\tiny\color{termgray},
  identifierstyle=\color{black},
  emph={self}, emphstyle=\color{termorange},
  numbers=left,
  numbersep=6pt,
  frame=single,
  rulecolor=\color{gray!50},
  tabsize=4,
  showstringspaces=false,
  breaklines=true,
  breakatwhitespace=true,
  columns=fullflexible
}
\newcommand{\Sf}{F}
\newcommand{\Sb}{B}
\newcommand{\Fr}{\text{Fr}}
\newcommand{\Cr}{\text{Cr}}

\newcommand{\Rc}{\mathcal R}
\newcommand{\Dc}{\mathcal D}
\newcommand{\Ac}{\mathcal A}
\newcommand{\Pc}{\mathcal P}
\newcommand{\Qt}{\mathcal Q}
\newcommand{\R}{\text{Re}}

\usepackage{lineno}

\begin{document}
\let\WriteBookmarks\relax
\def\floatpagepagefraction{1}
\def\textpagefraction{.001}
\shorttitle{openKARST: A novel open-source flow simulator for karst systems}
\shortauthors{Kordilla et~al.}

\title [mode = title]{openKARST: A novel open-source flow simulator for karst systems}                      

\author[1]{Jannes Kordilla}[type=editor,
                        orcid=0000-0003-4083-4491]
\cormark[1]
\ead{jannes.kordilla@idaea.csic.es}

\credit{Conceptualization, Methodology, Software, Validation, Formal analysis, Investigation, Writing - Original Draft, Visualization}

\author[1]{Marco Dentz}

\credit{Conceptualization, Methodology, Writing - Original Draft, Funding acquisition}

\author[1]{Juan J. Hidalgo}

\credit{Conceptualization, Methodology, Writing - Original Draft}

\affiliation[1]{organization={Institute of Environmental Assessment and Water Research (Institute of Environmental Assessment and Water Research, Spanish National Research Council (IDAEA-CSIC)), Groundwater and Hydrogeochemistry Group},
                addressline={Carrer de Jordi Girona 18-26}, 
                city={Barcelona},
                postcode={08034},
                country={Spain}}

\cortext[cor1]{Corresponding author}


\begin{abstract}
We introduce the open-source Python-based code \texttt{openKARST} for flow in karst conduit networks. Flow and transport in complex karst systems remain a challenging area of hydrogeological research due to the heterogeneous nature of conduit networks. Flow regimes in these systems are highly dynamic, with transitions from free-surface to fully pressurized and laminar to turbulent flow conditions and Reynolds numbers often exceeding one million. These transitions can occur simultaneously within a network, depending on conduit roughness properties and diameter distributions. \texttt{openKARST} solves the transient dynamic wave equation using an iterative scheme and is optimized through an efficient vectorized structure. Transitions from free-surface to pressurized flows in smooth and rough circular conduits are realized via a Preissmann slot approach in combination with an implementation of the Darcy--Weisbach and Manning equations to compute friction losses. To mitigate numerical fluctuations commonly encountered in the Colebrook--White equation, the dynamic switching from laminar to turbulent flows is modeled with a continuous Churchill formulation for the friction factor computation. \texttt{openKARST} supports common boundary conditions encountered in karst systems, as and includes functionalities for network import, export and visualization. The code is verified via comparison against several analytical solutions and validated against a laboratory experiment. Finally, we demonstrate the application of \texttt{openKARST} by simulating a synthetic recharge event in one of the largest explored karst networks, the Ox Bel Ha system in Mexico.
\end{abstract}


\maketitle

\doublespacing

\section{Introduction}

Karst aquifers are a significant water resource worldwide and, in many regions such as the Mediterranean, an indispensable source of freshwater~\citep{Chen2017a,Bresinsky2023a}. Given their importance for water resources
management, addressing and mitigating environmental impacts and risks is crucial. In contrast to aquifer systems in unconsolidated porous materials, karst systems are characterized by the presence of dissolutionally
enlarged conduits embedded within a fractured and porous matrix. While fracturing is common in many hard rock aquifers, the development of extensive conduit networks via chemical dissolution in both the phreatic and vadose
zones, sets karst environments apart from other hydrogeological systems. Conduit networks often reside in the hydraulically most dynamic parts, i.e.~close to the groundwater table, where the availability of dissolved
carbon dioxide enhances the dissolution process of limestone. However, geological processes such as subsidence, tectonic uplift or sea level changes can shift the position of dissolution horizons over time. As a result,
conduit networks may remain in the vadose zone under drained conditions or are fully submerged below or close to the groundwater table~\citep{Bakalowicz2015}.

Karst systems primarily receive water via diffuse infiltration through the porous matrix and via concentrated recharge, e.g.,~through sinkholes and strongly fractured pathways or fault zones. Infiltration commonly enters
via the epikarst and moves through the vadose and phreatic zones, which all can host rapid flow pathways embedded within the porous matrix~\citep{Kordilla2012,Shigorina2021}. This structure leads to pronounced contrasts in
hydraulic conductivity across the system~\citep{Schmidt2014}. Furthermore, despite their often thick vadose zones, karst aquifers are highly vulnerable to environmental impacts such as contamination from surface sources,
rapid transmission of pollutants in conduit systems~\citep{Neuman2005}, and changes in water quality and quantity due to shifts in land use and climate~\citep{Chen2018a}. Hence, this poses a challenge both for flow and
transport modeling tools that must account for wide array of flow and transport processes on various time scales~\citep{Jourde2023}.

Specifically, flow within conduit networks sets karst systems apart from many other hydrogeological systems. Conduits may reach diameters of several meters or more~\citep{Maqueda2023}, at which point they can also be
classified as caves, typically defined as conduits large enough for a human to enter.
Flow in karst conduit systems can cover the full spectrum ranging from laminar to turbulent and from free-surface to pressurized conditions, depending on local conduit geometry, pressure gradients, and boundary
conditions~\citep{Reimann2011,Shoemaker2008}. Flow may be initiated under dry conduit conditions as laminar free-surface flow, become turbulent once water depth increases, and potentially become pressurized when the water
level reaches the conduit ceiling. Due to these complexities, it remains a fundamental challenge in karst hydrodynamic modeling to accurately represent such transitions within spatially heterogeneous, three-dimensional
conduit networks~\citep{Jourde2023}.

Similar to the modeling of flow in fractured porous media~\citep{Berre2019}, for karst media one can distinguish equivalent porous media (EPM) or single continuum approaches, dual continuum (DC) models, combined discrete
continuum (CDC) models, and discrete conduit network (DCN) models~\citep{Kovacs2007,Hartmann2014,Jourde2023}.

The EPM approach~\citep{Scanlon2003} represents the hydraulically heterogeneous karst medium by an equivalent porous medium that is characterized by suitably defined averaged (upscaled) hydraulic
properties~\citep{Larocque1999}. The dual continuum approach models the karst system as two linearly interacting continua with very different hydraulic properties that are representative of the porous matrix and conduit
network~\citep{Cornaton2002,Kordilla2012}.

In the CDC approach, dominant conduits are explicitly modeled and embedded in a porous matrix~\citep{Kiraly1975}. A well-established implementation of this approach is MODFLOW-CFP~\citep{Shoemaker2007}, an extension of the
MODFLOW groundwater modeling suite developed by the U.S. Geological Survey (USGS). MODFLOW-CFP supports three distinct flow modes, referred to as CFP Mode 1, 2, and 3. Mode 1 simulates flow through a discrete conduit
network coupled to the MODFLOW matrix domain and allows both laminar and turbulent flows using the Hagen--Poiseuille and Darcy--Weisbach equations. Mode 2 does not include discrete conduits but modifies the porous medium
flow equations to mimic turbulent effects in highly permeable zones. Mode 3 combines Modes 1 and 2 to enable simultaneous representation of discrete conduits and EPM regions within the same model. MODFLOW-CFP also includes
exchange terms that represent dynamic interactions between the conduit system and the surrounding matrix. However, these tools assume steady or quasi-steady flow in conduits and are optimized for long-term,
matrix-dominated processes. They do not resolve the full transient dynamics of flow within the conduit network.

MODFLOW-USG~\citep{Kresic2018}, another MODFLOW-based tool, allows for unstructured grids and embedded features but supports only diffusive wave representations of open-channel flow and is primarily intended for surface
water or river simulations. It is not designed to handle fully pressurized, transient flow in submerged karst conduit networks.

Tools such as MODFLOW-CFP are developed to simulate long-term, catchment-scale dynamics where conduit--matrix exchange and regional hydraulic gradients dominate the flow behavior over timescales of years to decades. In
contrast, applications that require the simulation of short-term, high-frequency flow events such as flood waves, recharge pulses, or dynamic pressurization, models need to resolve fully transient conduit flow dynamics.
Some efforts have been made to adapt sewer system models such as SWMM~\citep{SWMM2017} to simulate karst conduits~\citep{Campbell2002}, and to represent recharge, storage, and inter-compartmental exchange processes via
coupling with external hydrological or reservoir models~\citep{Chen2014,Gabrovsek2018}. However, SWMM itself was originally developed for sewage drainage networks rather than for hydrogeological applications. Therefore, it
lacks the flexibility to represent karst-specific features such as the explicit use of the Darcy--Weisbach equation for free-surface flows, or physically based correction terms in the momentum equation that account for
spatially distributed recharge. 

In the context of groundwater modeling, other efforts were made to solve the Saint-Venant equations. The MODBRANCH model~\citep{Swain1996} simulates transient one-dimensional stream--aquifer interactions and was extended
by~\citet{ZhangLerner2000} to enable simulation of adit systems via the addition of a Preissmann slot and dynamic pressurization. Following these works,~\citet{Reimann2011} developed ModBraC, which extends the approach to
solve the Saint-Venant equation under variably saturated conditions. These studies represent important steps towards integrating transient conduit hydraulics into groundwater models. Their primary focus was on surface
water--groundwater interactions or matrix--conduit exchange within the MODFLOW framework. 

In this work, we introduce \texttt{openKARST}, a lightweight Python-based flow simulator for fully transient conduit flow dynamics in complex karst networks. In contrast to previous dual-domain implementations,
\texttt{openKARST} is centered on complex conduit networks, and independent of a porous matrix model. The model solves the full one-dimensional Saint-Venant (dynamic wave) equations and accounts for both laminar and
turbulent flow regimes, including dynamic transitions between free-surface and pressurized conditions in heterogeneous conduit geometries. Accurate modeling of such dynamics is essential for the simulation of flood waves,
recharge pulses, and drought responses under rapidly changing boundary conditions. To offer a flexible, transparent, and accessible development platform, \texttt{openKARST} is implemented entirely in Python. The
graph-based, grid-free structure allows to represent synthetic and real-world karst networks, while the modular design supports community-driven extension and integration with modern characterization and stochastic
modeling approaches~\citep{Collon2017,Maqueda2023}.

The paper is structured as follows.  \ref{sec:methdology} presents the overall methodology. It states the network scale flow problem in terms of the continuity and momentum conservation equations that constitute the dynamic wave or Saint-Venant equations. It discusses the space and time discretization of the flow equations including upstream weighting and inertial damping, and the iterative solution of the implicit non-linear system of equations by Picard iteration. Then it presents the geometrical and physical relations to close the resulting system of discrete equations, as well as boundary and initial conditions. More detailed derivations are provided in  \ref{app:numerics}.
 \ref{sec:verification} discusses the verification of the code in comparison to analytical solutions for steady-state free-surface and pressurized flow, validation against experimental data, and an application example for flow in a real karst network geometry.
In  \ref{sec:demonstration} we demonstrate the application of \texttt{openKARST} to a real-world network by simulating transient recharge in the Ox Bel Ha cave system in Mexico. This example highlights the scalability and ability of the model to capture dynamic flow responses in large, complex conduit networks.
Finally, in  \ref{sec:conclusion} we summarize the main findings and discuss future directions, including possible extensions of \texttt{openKARST} to coupled matrix--conduit systems and other hydrogeological processes.

\section{Methodology}
\label{sec:methdology}
\begin{figure}
\centering
\caption{Spatial discretization of the underlying network showing both network nodes (red circles) and computational nodes (black dots). Note that the spacing $\Delta x$ can be variable and different for each conduit of length $L$.\label{fig:network_discretization}}
\includegraphics[width=0.5\linewidth]{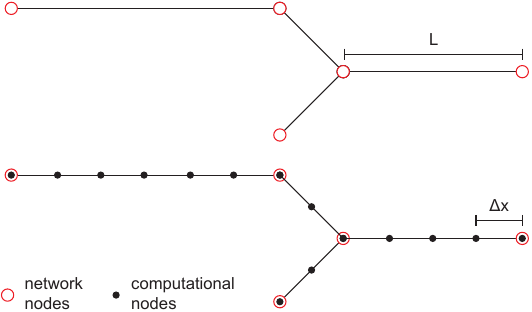}
\end{figure}

 We consider flow through a karst system that consists of a network of conduits
as illustrated in  \ref{fig:network_discretization}. Flow along single conduits is
quantified by the Saint-Venant equations, which  describe the average flow
velocity or flow rate along the conduit. At the network nodes, mass is
conserved. In the following, we first summarize the Saint-Venant equations for
flow along a single conduit and their implementation on the karst network. Then
we describe the spatial and temporal discretization of the governing equations
and their numerical solution. A description of the user interaction, including pre- and post-processing options, visualization and a minimal usage example, is provided in  \ref{app:user}.

\subsection{Network flow}

 The karst system is represented as a network of $N_\circ$ nodes and $N_c$
conduits, see  \ref{fig:network_discretization}. Each node is connected to
$d$ conduits, which defines the degree of the node. Conduits are characterized
by the length $L$ [L] and a constant hydraulic diameter $\Dc = 4\Ac/\Pc$ [L], where
$\Ac$ [L$^2$] is the cross-sectional area and $\Pc$ [L] the
perimeter of the conduit. We define the equivalent conduit radius by $\Rc =
\Dc/2$. We consider transient free-surface and pressurized flow, which is quantified
by the flow rate or discharge $Q$ [L$^3$/T] and water depth $y$ [L] along the conduit. From the water depth
and the underlying conduit geometry, one can determine the
cross-sectional width of the free-surface $W$, the discharge area $A \leq
\Ac$, and the wetted perimeter $P$ [L]. The wetted hydraulic radius and diameter
are then given by $R = A/P$ [L] and $D = 4A/P$ [L]. Note that $D = 4R$ by definition. The average
flow velocity along the conduit is given by $v = Q/A$ [L/T]. The areas and lengths characterizing the water
phase along a conduit are functions of the water depth $y$ only and can be
determined based on the conduit shape. Conduits are represented as
one-dimensional objects. The flow behavior is characterized by the flow rate
$Q$ and water depth $y$.

Mass is conserved in the system and fluid density is constant. While storage is physically associated with the conduit volume, a control volume around each node is defined to formulate the local mass balance. The temporal change of the nodal volume $V_i$ [L$^3$] at node $i$ is governed by the net inflow from all connected conduits and external recharge or discharge:
\begin{equation}
\label{eq:nodal_volume1}
\frac{d V_{i}}{d t} = \sum_{l=1}^{d_i} Q_{i,l} + \Qt_{i,r}\,.
\end{equation}

Here, $Q_{i,l}$ denotes the flow rate from conduit $l$ into node $i$, and $\Qt_{i,r}$ represents any prescribed recharge or extraction rate at node $i$. Therefore, the nodal volume $V_i$ represents the change in stored water around a node, based on the adjacent conduit geometry.
In the following, the subscript $i$ counts the nodes, the subscript $l$ the conduits.

Flow in a conduit $l$ connecting the nodes $i_1$ and $i_2$ is determined by the
Saint-Venant equations~\citep{Saintvenant1871}, which describe transient free
surface flow through pipes or channels with variable geometry. They quantify the
cross-sectionally averaged flow along conduits and are based on the following
assumptions. (1) Gentle conduit bed slope: The slope of the conduit bottom is assumed to be smaller than 10\%. (2)
Hydrostatic pressure distribution: The vertical acceleration of the flow is
negligible, that is, the pressure distribution at any cross-section is
hydrostatic. (3) Boundary friction: Friction at the boundary is modeled similar
to steady-state flows, based on a constitutive relationship between flow
velocity and shear stress. Under these assumptions, conservation of mass and
momentum along the conduit are governed by the following equations~\citep{Saintvenant1871,Chow1959}:
\begin{align}\label{eq:continuity}
&\frac{\partial A}{\partial t} + \frac{\partial Q}{\partial x}  = q
\\
\label{eq:momentum}
& \frac{\partial Q}{\partial t} + \frac{\partial}{\partial
  x}\left(\frac{Q^2}{A}\right) + g A \frac{\partial H}{\partial x} + g A \Sf = 0,
\end{align}
where $t$ is time [T], $x$ distance [L], $g$ gravitational
acceleration [L T$^{-2}$], $\Sf$ friction slope [-], $H$
hydraulic head [L], and $q$ [L$^2$T$^{-1}$] an areal recharge or discharge flux. The friction
slope $\Sf$ expresses the action of the wall shear stress on fluid motion and is specified in
 \ref{sec:friction}. It depends on the hydraulic radius $R$, which in
turn is a function of the water depth $y$. Under free-surface flow, we employ the Manning
formula, and under pressurized conditions the formula corresponding to the
Darcy--Weisbach equation. The hydraulic head $H$ can be decomposed into the water depth
$y$ and the bottom elevation $z$ [L] as $H = z + y$, and the flow rate $Q$ can be written as the product of flow velocity $v$ times cross-sectional area $A$ as $Q = v A$. The cross-sectional area $A$ is a function of the water depth $y$. Its
functional dependence on the shape of the conduit cross-section as
discussed in  \ref{sec:surfaces}.  
With these definitions, Eqs.~\eqref{eq:continuity} and ~\eqref{eq:momentum} can be combined into (see  \ref{app:derivation})
\begin{equation}
\label{eq:momentum2}
\frac{\partial Q}{\partial t} = 2v \left(\frac{\partial A}{\partial t} - q\right)
+
v^2 \frac{\partial A}{\partial x} - gA \frac{\partial y}{\partial x} - gA \Sf +
gA \Sb,
\end{equation}
where $\Sb = -d z/dx$ is the conduit bed slope [-]. The water depths at
either end of the conduit are given by the nodal water depths $y_{i_1}$ and
$y_{i_2}$, where $i_2$ denotes the downstream and $i_1$ the upstream
node. The first assumption on gentle bed slopes may be violated locally in realistic karst formations. In the case of steep bed slopes, the Saint-Venant equations can be modified as discussed in~\citet{Ni2019}. In this
paper, we use the Saint-Venant equations, which provide a powerful framework to quantify transient conduit flow with moderate bed slopes as long as vertical acceleration can be disregarded. 

Conduit flow can be characterized by the Froude number~\citep{Chow1959}
\begin{equation}
\label{eq:Fr}
\Fr = \frac{\vert v\vert }{\sqrt{g \frac{A}{W}}},
\end{equation}
which compares the impact of inertia and gravity on the flow behavior. For $\Fr < 1$, gravity dominated and flow is subcritical. A
perturbation of the water height propagates both in the upstream and downstream
direction. For $\Fr = 1$, flow is critical, a perturbation remains at the same
location. For $\Fr > 1$ flow is supercritical and a flow perturbation moves only
downstream.

The flow system is determined by \eqref{eq:continuity} and ~\eqref{eq:momentum2} where the discharge $Q$ and the water depth $y$ are the
dependent variables and functions of distance and time. The cross-sectional area
for flow $A$ is a time-dependent and geometry-specific property depending on the
water depth. Under pressurized conditions, the cross sectional area $A$ and the
flow rate $Q$ are constant. Thus, the mass conservation statement given by
Eq.~\eqref{eq:continuity} is trivially fulfilled, and Eq.~\eqref{eq:momentum2}
reduces to
\begin{equation}
\label{eq:DWmomentum}
     - \frac{\partial y}{\partial x} + \Sb = F,
\end{equation}
the Darcy--Weisbach equation in the head formulation.

\subsection{Space discretization}

 A single conduit is discretized into segments of length $\Delta x$ separated by computational nodes. Between the computational nodes along each segment, uniform properties (e.g.,~diameter, roughness, shape) are assumed. If spatial heterogeneity is required, the conduit can be discretized into smaller segments with varying parameters. In order not to complicate notation, we consider the computational nodes as nodes in the network and the segments as network conduits, but keep in mind that there is a distinction between the network nodes and the computational nodes. In fact, the coordination number, that is, number of conduits connected to a node is determined by the network topology for the network nodes and is equal to $2$ for the computational nodes within a conduit. In the following, all nodes, computational and network nodes alike, are counted by the index $i$ and conduits by the index $l$. Water depths $y_i$ are evaluated at nodes, flow rates $Q_l$ at conduits. Quantities that belong to a conduit $l$ are denoted by the subscript $l$, for example, the discharge surface $A_l$ and the wetted radius $R_l$. Conduit attributes that contribute to or are evaluated at node $i$ are denoted by the subscripts $(i,l)$. For example, $Q_{i,l}$ denotes the contribution of conduit $l$ to the flow rate at node $i$, and $A_{i,l}$ denotes the discharge area of conduit $l$ evaluated using the water depth $y_i$ at node $i$. Properties characterizing the geometry of the water phase at the end of the conduit, such as hydraulic radius $R_{i,l}$, width of the free water surface $W_{i,l}$, and discharge area $A_{i,l}$ are evaluated in terms of the water depth $y_i$ at the nodes such that,
\begin{equation}
A_{i,l} = A_l(y_i), \qquad R_{i,l} = R_l(y_i), \qquad W_{i,l} = W_l(y_i).
\end{equation}
The functional forms of $A(y)$, $R(y)$, and $W(y)$ for circular
and rectangular conduit cross-sections are given in  \ref{sec:surfaces}.
Conduit properties that are evaluated at the center of the conduit are
determined in terms of the average water depth,
\begin{equation}
\bar y_l = \frac{y_{i_1} + y_{i_2}}{2},
\end{equation}
and denoted by an overbar, that is,
\begin{equation}
\bar A_{l} = A_l(\bar y_l), \qquad \bar R_{l} = R(\bar y_l), \qquad \bar W_{l}
= W(\bar y_l). 
\end{equation}
In order to improve numerical stability, we consider in the following also the upstream weighted averages of these quantities. Upstream weighted 
 quantities are marked by the subscripts $(\alpha,l)$. For example, the upstream weighted discharge area for conduit $l$ is defined as
\begin{equation}
\bar A_{\alpha,l} = A_{i_1,l} + \alpha(\bar A_l - A_{i_1,l}),
\end{equation}
where $\bar A_l$ is the discharge area evaluated at $\bar y_l$.
For $\alpha = 1$, $\bar A_{1,l} \equiv \bar A_l$. The same notation conventions are employed
for the quantities derived from the conduit properties such as the
velocity $\bar v_l = Q_l \bar A_l$ and the friction slope $\bar F_{\alpha,l}$ which is
determined from the upstream-weighted hydraulic radius $\bar R_{\alpha,l}$. 

With these conventions, we discretize the continuity
equation~\eqref{eq:continuity} inside the conduit as
\begin{equation}
\label{eq:cont_cond}
\frac{d V_i}{dt} = \sum_{l = 1}^2 Q_{i,l} + \Qt_{i,r},
\end{equation}
where $V_i = \Delta x A_i$ is the volume of water at node $i$ and $\Qt_{i,r}$ the nodal recharge or discharge rate. Eq.~\eqref{eq:cont_cond} has the same form as
Eq.~\eqref{eq:nodal_volume1}. Flow rates are defined in the conduits, water heads
at the nodes.
At network and computational nodes, the nodal volume $V_i$ is defined as a function of water head $y_i$ [$L$] and thus,
\begin{equation}
  \label{eq:nodal_volume2}
  S_{i} \frac{d y_{i}}{d t} = \sum_{l=1}^{d_i} Q_{i,l} + \Qt_{i,r}, \qquad S_i = \frac{d V_i}{d y_i},
\end{equation}
where $S_i$ denotes the effective free-surface area associated with node $i$. This surface area is computed as the sum of contributions from all conduits $l$ connected to the node
\begin{equation}
  \label{eq:nodal_area}
  S_{i} = \sum_{l = 1}^{d_i} S_{i,l}, \qquad S_{i,l} = \frac{W_{i,l} + \bar{W}_l}{2} \cdot \frac{\Delta x}{2}\,.
\end{equation}
Each conduit contributes half of its segment length to the control volume around a node. The nodal volume therefore approximates the storage compartment formed by all contributions of adjacent half-length conduits, evaluated based on the local water depth $y_i$. Source and sink terms (e.g.,~recharge or extraction) are defined at nodes and distributed along the conduit via interpolation. The recharge volume at node $i$ is given by
\begin{equation}
\Qt_{i,r} = q_i \, d_i \, \frac{\Delta x}{2}, \qquad \bar{q}_l = \frac{q_{i_1} + q_{i_2}}{2} = \frac{\Qt_{i_1,r}}{d_{i_1} \Delta x} + \frac{\Qt_{i_2,r}}{d_{i_2} \Delta x}\,.
\end{equation}
This formulation allows recharge to influence both the nodal storage and the conduit momentum balance through appropriate interpolation of fluxes.

For the spatial discretization of the momentum equation~\eqref{eq:momentum2}, upstream weighting is applied for the pressure term and friction slope as outlined below. Furthermore, in order to improve the stability of the
numerical solution, inertial damping is employed~\citep{Fread1996}. This gives,
\fontsize{7.5pt}{10.5pt}\selectfont\begin{equation}
\label{eq:momentum3}
\frac{d Q_l}{dt} = 2 \alpha \bar v_{l} \left(\frac{d \bar A_{l}}{dt} - q_l\right) + \alpha \bar v_{l}^2
\frac{A_{i_2,l} - A_{i_1,l}}{\Delta x} - g \bar A_{\alpha,l} \frac{y_{i_2} -
  y_{i_1}}{\Delta x} - g \bar A_{\alpha,l} \bar \Sf_{\alpha,l} + g \bar
A_{l} \bar \Sb_l. 
\end{equation}\normalsize
Recall that $i_2$ denotes the downstream and $i_1$ the upstream nodes. Both upstream weighting and
inertial damping are determined by the parameter $\alpha$, which depends on the
local Froude number as outlined in the following.  With this space
discretization, flow in the network is described by Eqs.~\eqref{eq:nodal_volume2} and ~\eqref{eq:momentum3} supplemented by
constitutive relations for the wetted discharge area $A_{i,l}$, free-surface area
$S_{i,l}$, and friction slope $F_l$, detailed in  \ref{app:closures}.  
 
\subsubsection*{Upstream weighting and inertial damping}
\label{subsec:upstream_weighting}
 In order to improve numerical stability, upstream weighting and inertial
damping~\citep{Fread1996,SWMM2017}
are implemented in the spatially discrete momentum
equation~\eqref{eq:momentum3}. The value of $\alpha$, which determines both the
upstream weighting of the pressure term and friction slope as well as the damping 
depends on the value of the local Froude number, which is defined in terms of
the average flow velocity $\vert \bar v_l\vert $, discharge area $\bar A_l$ and free
surface width $\bar W_l$ as
\begin{equation}
    \Fr_l = \frac{\vert \bar{v_l}\vert }{\sqrt{g\bar{A_l}/\bar{W_l}}},
\end{equation}
 Recall that for $\Fr_l<1$ the flow is in a sub-critical state, that is,
gravitational forces dominate and flow is stable. For $\Fr_l>1$ the flow is in a
supercritical state dominated by inertial forces and surface waves move in the
direction of the flow gradient. In this case, the average discharge area and
average hydraulic radius are upstream weighted. The weight $\alpha$ is set to
one for subcritical conditions and zero for supercritical. It interpolates
linearly between one for $\Fr_l = 0.5$ and zero for $\Fr_l = 1$, that is,
\begin{equation}
\alpha = 
\begin{cases} 
1 & \text{for } \Fr_l \leq 0.5 \\
2(1 - \Fr_l) & \text{for } 0.5 < \Fr_l < 1 \\
0 & \text{for } \Fr_l \geq 1 
.\end{cases}\end{equation}
Under pressurized flow conditions $\alpha$ is set to zero, that is, no upstream
weighting occurs and inertia terms are suppressed in line with the
Darcy--Weisbach equation for pressurized conduit flow.

\subsection{Time discretization}

 The continuity equation~\eqref{eq:nodal_volume2} is discretized in time
using a Crank--Nicholson scheme
\begin{equation}
  \label{eq:nodal_volume_discrete}
    y_{i}^{k+1} = y_{i}^{k} + \frac{\Delta t}{S_{i}^{k+1}}
    \frac{1}{2}\left(\sum_{l=1}^{d_{i}} Q^{k+1}_{i,l} + \Qt^{k+1}_{i,r} + \sum_{l=1}^{d_{i}}
    Q^{k}_{i,l} + \Qt^{k}_{i,r}\right). 
\end{equation}
The indices $k$ in the following count the time steps. The momentum
conservation equation~\eqref{eq:momentum3} is discretized using a backward Euler scheme,
\begin{align}
\frac{Q_{l}^{k+1} - Q_{l}^{k}}{\Delta t_k} &=
   2\alpha\bar v_{l}^{k+1} \left(\frac{\bar A_{l}^{k+1} -
     \bar{A}_{l}^{k}}{\Delta t_k} - q_l^{k+1} \right)
   +
   \alpha \left(\bar v_{l}^{k+1}\right)^2 \frac{A_{i_2,l}^{k+1}
   -
   A_{i_1,l}^{k+1}}{\Delta x} 
\nonumber
\\   
&   -
g\bar{A}^{k+1}_{\alpha,l} \frac{y^{k+1}_{i_2} - y^{k+1}_{i_1}}{\Delta x}
- g \bar{A}^{k+1}_{\alpha,l} \bar \Sf_{\alpha,l}^{k+1} + g
\bar{A}_{l}^{k+1} \bar \Sb_{l}, 
\label{eq:momentum_discrete}
\end{align}
where the time increment $\Delta t_k$ may vary with the time step $k$ as discussed in  \ref{app:time_step}.

In this discretization scheme, pressurized flow conditions are not well defined. A node becomes pressurized if the discharge area $A_i$ reaches the full cross-sectional area of the conduit, and hence the free
surface area contribution vanishes. Physically, this means that the water depth $y_i$ reaches the conduit ceiling, beyond which it cannot increase further under the standard free-surface formulation. Therefore, once
a node is pressurized, the water depth $y_i$ cannot be updated anymore using Eq.~\eqref{eq:nodal_volume_discrete}. In order to circumvent this difficulty, small but finite free-surface areas $S_{i,l}$ are
assigned to pressurized conduits using the concept of the Preissmann slot~\citep{Cunge1964}. Details are given in  \ref{subsec:pressurized}.

Analogous to the challenges of modeling pressurized conduits, the drying and
rewetting dynamics also require numerical approximations to effectively address
limitations of the governing equations when water depths become zero or
negative. When this occurs, Eq.~\eqref{eq:momentum_discrete} cannot be solved
because then the hydraulic radius $R = 0$ and thus the friction slope $\Sf$ are not
defined. Therefore we set a global lower limit for the water depth of $y_{min} =
\SI{1e-12}{m}$, which sustains a finite but negligible flow component within conduits.

\subsection{Picard iteration}

 Eqs.~\eqref{eq:nodal_volume_discrete} and ~\eqref{eq:momentum_discrete} form
a coupled non-linear system, which can be written as
\begin{align}
\mathbf y^{k+1} &= \mathbf F(\mathbf y^k, \mathbf y^{k+1},\mathbf Q^k,\mathbf Q^{k+1})
\\
\mathbf Q^{k+1} &= \mathbf G(\mathbf y^k, \mathbf y^{k+1},\mathbf Q^k,\mathbf Q^{k+1}).
\end{align}
We define the vectors $\mathbf y = (y_1,\dots,y_{N_\circ})^\top$ and $\mathbf Q
= (Q_1,\dots,Q_{N_c})^\top$, where the subscript $\top$ denotes the
transpose. The system is solved using the relaxed Picard iteration~\citep{Langtangen2017}. At each iteration step, we obtain
\begin{align}
\hat{{\mathbf Q}}^{j+1} &= \mathbf G(\mathbf y^{k},\mathbf y^{k,j},\mathbf Q^k, \mathbf{ Q}^{k,j}) 
\\
\mathbf Q^{k,j+1} &= \omega \hat{{\mathbf Q}}^{j+1} + (1 - \omega) \mathbf Q^{k,j}
\\
\hat{{\mathbf y}}^{k,j+1} &= \mathbf F(\mathbf y^k,\mathbf y^{k,j},\mathbf Q^k,\mathbf Q^{k,j+1})
\\
\mathbf y^{k,j+1} &= \omega \hat{{\mathbf y}}^{j+1} + (1 - \omega) \mathbf y^{k,j}
\end{align}
for the initial values $\mathbf y^{k,j = 0} = \mathbf y^{k}$ and $\mathbf Q^{k,j = 0} = \mathbf Q^{k}$. The index $j$ counts the Picard iterations. The variables with a hat refer to the unrelaxed quantities that are obtained directly from the values of the previous iteration. The relaxation factor $0< \omega \leq 1$ is introduced to ensure convergence. It prevents the iterated value from being too different from the values at the previous iteration. The value of $\omega$ is set to 0.8 by default.  

Nodes with $y_i^{k,j+1} < y_{min}$ are considered dry. To avoid
outflow from dry nodes we check after each Picard iteration if the nodes at
either end of a conduit satisfy $y^{k,j+1}_i \le y_{min}$. If this is the case,
the flow rate of that conduit is set to $Q^{k,j+1}_{l} = Q_{min} =
\SI{1e-12}{m^3 s^{-1}}$ keeping the flow direction, and the water depth is set
$y^{k,j+1}_i = y_{min}$, which prevents negative water depths. 

The iteration stops when the convergence criterion
in terms of subsequent water heads is reached. That is, convergence is reached
when the absolute differences between all the elements of the current water depths vector,
$\mathbf{y}^{k,j+1}$, and the corresponding elements of the previous water
depths vector, $\mathbf{y}^{k,j}$, are smaller than a tolerance $\delta$,
\begin{equation}
    \text{Convergence} \quad \text{if} \quad \forall i, \quad \left\vert   y_i^{k+1,j+1} - y_i^{k+1,j} \right\vert   < \delta\,.
\end{equation}
The maximum number of Picard iterations is set to $20$ by default, and the
tolerance to $\delta = 10^{-8}$. If convergence is reached, $\mathbf Q^{k+1} =
\mathbf Q^{k,j+1}$ and $\mathbf y^{k+1} = \mathbf y^{k,j+1}$. The Picard
iteration for the next time step starts.

\subsection{Closure Relations}
\label{sec:constitutive}
 The numerical solution of Eqs.~\eqref{eq:nodal_volume_discrete} and ~\eqref{eq:momentum_discrete} requires the definition of a series of geometrical closure relations. Specifically, we need expressions for the discharge area $A$, the width $W$ of the free water surface, and the wetted perimeter $P$ as functions of the local water depth, which depend on the conduit geometry. These relations are essential to link the hydraulic state variables to the conduit cross-section and complete the flow formulation. We assume here that the conduits have a circular cross section with constant diameter $\Dc$. Different conduit shapes can be implemented in a straightforward manner. Furthermore, we need to define the Preissmann slot for the modeling of pressurized flow conditions as well as the formulae for the friction slope $\Sf$. The closure relation are detailed in  \ref{app:closures}.

\subsection{Boundary and initial conditions}

   Head boundary conditions at inflow or outflow nodes are enforced by
setting the water depth $y^{k,j+1}_i$ equal to the corresponding fixed water
depth $y_{i,f}$ after each Picard iteration
\begin{equation}
y^{k,j+1}_i = y_{i,f}.
\end{equation}
Flow-rate boundary conditions at inflow or outflow nodes are imposed by prescribing the total nodal discharge, which is defined by
\begin{equation}
\Qt_i = \sum_{l=1}^{d_i} Q_{i,l}.  
\end{equation}
Using this notation, the continuity equation~\eqref{eq:nodal_volume_discrete}
reads as
\begin{equation}
 y_{i}^{k+1} = y_{i}^{k} + \frac{\Delta t}{S_{i}^{k+1}}
    \frac{1}{2}\left(\Qt_i^{k+1} + \Qt_i^{k}\right). 
\end{equation}
A prescribed volumetric inflow value $\Qt_{i,f}$ at a boundary node $i$ is enforced by setting
$\Qt_{i}^{k,j+1}$ at each Picard iteration equal to
\begin{equation}
\Qt_i^{k,j+1} = \sum_{l=1}^{d_{i}} Q^{k,j+1}_{i,l} + \Qt_{i,f}.  
\end{equation}
This implies that under steady-state conditions the sum of discharges in the
conduits connected to node $i$ is $\Qt_i = -\Qt_{i,f}$.

   Initial conditions must be set for the water depth $y_i$ at the nodes and the flow
rate $Q_l$ in the conduits. By default, all values are initially set to zero.

\section{Verification and validation}
\label{sec:verification}
 
In the following subsections we first verify the developed model through a
series of analytical tests. We consider steady-state conditions, different
geometries and boundary conditions, as well as free-surface and pressurized
flows. For the first three cases we conduct a convergence study with respect to
the resolution $\Delta x$ using the percentage root mean square error (RMSE) as an
accuracy metric. Finally we compare the code to a laboratory experiment in order
to validate the correct implementation of the full dynamic wave equation under
transient conditions. Equilibrium conditions for the steady-state cases at $t=\SI{5000}{s}$ are
computed with a time increment $\Delta t=\SI{0.1}{s}$. All cases used an
under-relaxation of $\omega=0.8$. Initial conditions are $y=\SI{0}{m}$ and
$Q=\SI{0}{m^3 s^{-1}}$ unless otherwise specified. The water density is set to
$\rho=\SI{1000}{kg m^{-3}}$ and the dynamic viscosity is $\mu = \SI{0.001}{kg
  m^{-1}s^{-1}}$.
 
\subsection{Steady-state free-surface channel flow}
\label{sec:recharge}
 In this section, we verify the numerical implementation for free
surface flow in a rectangular channel under steady-state
conditions. The governing Eqs.~\eqref{eq:continuity} and ~\eqref{eq:momentum2} then reduce to
\begin{align}\label{eq:continuity_ex}
&\frac{\partial Q}{\partial x}  = q
\\
\label{eq:momentum_ex}
&  \frac{Q^2}{gA^3}\frac{\partial A}{\partial x} - \frac{\partial
  y}{\partial x} - \frac{2Q q}{A} - \Sf +
 \Sb = 0,
\end{align}
where we used that $Q = v A$. Eq.~\eqref{eq:momentum_ex} provides an
equation for the bedslope $\Sb$ in terms of a given water depth, which can be
written as
\begin{equation}
\label{eq:zx}
\frac{dz}{dx} = \left(\frac{Q^2}{gA^3} \frac{d A}{dy}  - 1 \right)
\frac{\partial y}{\partial x} - \frac{2Q q}{g A^2} - \Sf. 
\end{equation}
Note that $Q = \Qt_f + q x$, where $\Qt_f$ is the prescribed flow rate at the left boundary.  
The rectangular channel has width $b$ such that the discharge area and the
hydraulic radius are
\begin{equation}
A(y) = b y, \qquad R(y) = \frac{by}{b+2 y}.
\end{equation}
We define the flow rate and recharge rate per channel width as $Q' = Q/b$ and $r_0 = q/b$. Thus, we can write Eq.~\eqref{eq:zx} as
\begin{equation}\label{eq:zx2}
\frac{dz}{dx} = \left(\frac{Q^{\prime 2}}{gy^3}  - 1 \right)
\frac{\partial y}{\partial x} - \frac{2Q' r_0}{g y^2} - \Sf, \qquad Q' = Q'_f + r_0 x. 
\end{equation}
For a broad channel with $b \gg y$, we can set
$R = y$ and the Manning friction slope given by Eq.~\eqref{eq:manning_sf} becomes
\begin{equation}
    \Sf(y) = \frac{n^2 Q' \vert Q'\vert }{y^{\frac{10}{3}}}.
\end{equation}
Following~\citet{Delestre2013}, we prescribe the water depth $y_0(x)$,
from which we can obtain the bottom slope
$z(x)$ by integration of Eq.~\eqref{eq:zx} from $0$ to $x$ as
\begin{equation}
z(x) = \int_0^x dx' \left\{\left(\frac{Q^{\prime 2}}{g y_0(x')^3}  - 1 \right) \frac{\partial y_0(x')}{\partial
  x'} - \frac{2Q' r_0}{g y_0(x')^2} - \Sf[y_0(x')] \right\}. 
\end{equation}

In the following, we numerically solve the Saint-Venant equations
\begin{align}
&\frac{\partial y}{\partial t} + \frac{\partial Q'}{\partial x}  = 0
\\
&\frac{\partial Q'}{\partial t} = 2v \left(\frac{\partial y}{\partial
  t} -r_0 \right) +
v^2\frac{\partial y}{\partial x} - gy \frac{\partial y}{\partial x} - gy \Sf +
gy \Sb
\end{align}
with the bedslope
\begin{equation}
\Sb = - \left(\frac{Q^{\prime 2}}{g y_0(x)^3}  - 1 \right) \frac{\partial y_0(x)}{\partial
  x} + \frac{2Q' r_0}{g y_0(x)^2} + \Sf[y_0(x)]
\end{equation}
for a given $y_0(x)$ as specified below. The numerical steady-state solution is
then compared to the exact $y_0(x)$. 

\begin{figure*}
\caption{Comparison of the numerical solution for the water depth
      at steady-state with the analytical solution
      Eq.~\eqref{eq:analyical_delestre1} for different resolutions
      $\Delta x=\SIlist[list-units=single]{1;5;10;25;50}{m}$. Notice that
      the channel base (lightgray) is only shown for a resolution of
      $\Delta x=\SI{1}{m}$. Errors are shown for all resolutions with
      lighter gray for higher values of $\Delta x$. RMSE values are computed
      based on the percentage errors.\label{fig:delestre_3_3_1_2}}
\includegraphics[width=1\linewidth]{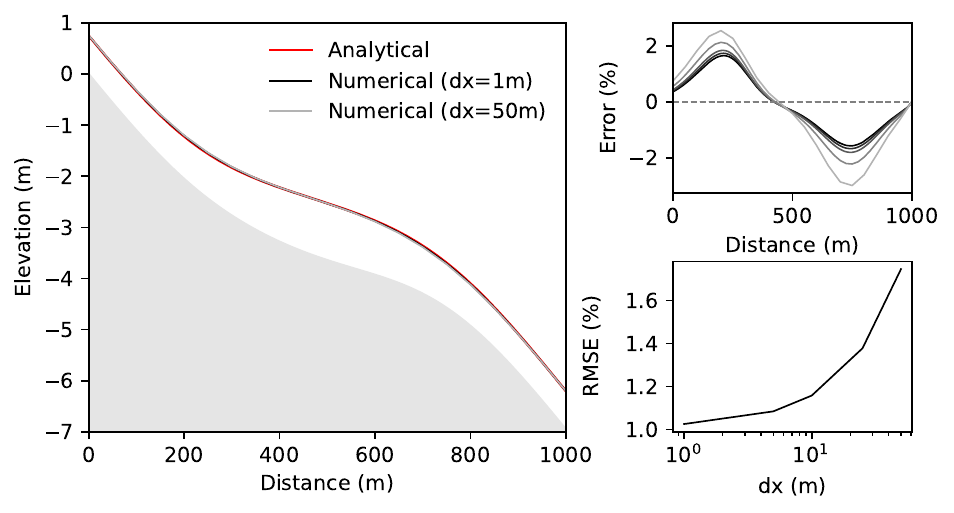}
\end{figure*}
 
\subsubsection{Gaussian profile}

 In this verification example, we consider free-surface flow without
recharge, that is, $r_0 = 0$, in a channel of length $L =
1000$ m with the prescribed equilibrium profile given by the
Gaussian-shaped function~\citep{Delestre2013}
\begin{equation}
\label{eq:analyical_delestre1}
y_{0}(x)= \left( \frac{4}{g} \right)^{\frac{1}{3}} \left( 1 +
   \frac{1}{2} \exp \left[ -16 \left( \frac{x}{1000} - \frac{1}{2}
   \right)^2 \right] \right). 
\end{equation}
At the left boundary at $x=\SI{0}{m}$ we specify the flow rate
$\Qt'_f=\SI{2.0}{m^2 s^{-1}}$. At the right boundary we set the
constant water depth $y_f = y_{0}(\SI{1000}{m}) = \SI{0.7483}{m}$.
We consider channel discretizations of length $\Delta
x=\SIlist[list-units=single]{1;5;10;25;50}{m}$. The Manning roughness
coefficient is set to $n=\SI{0.033}{m^{-1/3}s}$. The initial height
distribution is set to $y(x,t = 0) = \SI{0}{m}$. The numerical simulations
converge to steady-state after the time $t = \SI{5000}{s}$. 

The comparison of the numerical and the analytical solution is shown in
\ref{fig:delestre_3_3_1_2}. Its demonstrates good agreement and
a maximum error of about \SI{\pm 1.8}{\percent} for the highest
resolution of $\Delta x=\SI{1}{m}$ near the steepest parts of the
channel. For the lowest resolution with $\Delta x=\SI{50}{m}$ the results are still
acceptable with an error of about \SI{\pm 2.5}{\percent}. The
percentage RMSE converges towards an error of \SI{\pm 1}{\percent} at
the highest resolution and increases to about \SI{\pm 1.7}{\percent}
for the lowest resolution.
 
\subsubsection{Wavy profile}

\begin{figure*}
\caption{Comparison of the numerical solution for the water depth
      at steady-state with the analytical solution
      Eq.~\eqref{eq:analyical_delestre2} for different resolutions
      $\Delta
      x=\SIlist[list-units=single]{1;5;10;50;100;200}{m}$. Notice that
      the channel base (lightgray) is only shown for a resolution of
      $\Delta x=\SI{1}{m}$. Errors are shown for all resolutions with
      lighter gray for higher values of $\Delta x$. RMSE values are computed
      based on the percentage errors.\label{fig:delestre_3_3_1_9}}
\includegraphics[width=1\linewidth]{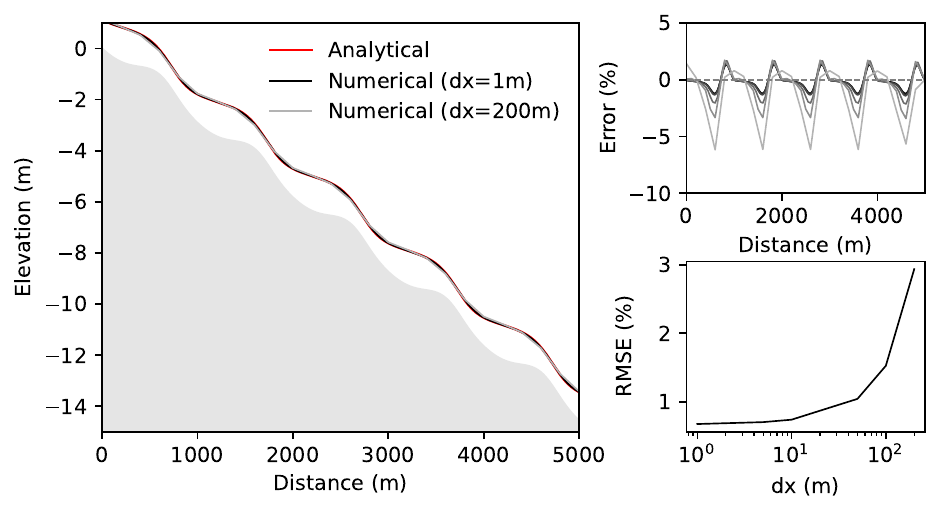}
\end{figure*}

 Next we consider a verification example of free-surface flow in a channel of length
$L=\SI{5000}{m}$ for $r_0 = 0$ for the sinusoidal steady-state profile
\begin{equation}
\label{eq:analyical_delestre2}
y_{0}(x) = \frac{9}{8} + \frac{1}{4} \sin\left(\frac{10 \pi x}{L}\right)
\end{equation}
At the left boundary at $x=\SI{0}{m}$, we set the constant inflow
$Q_f=\SI{2.0}{m^2 s^{-1}}$. At the right boundary at
$x=\SI{5000}{m}$ we specify the constant depth
$y_f = y_{0}(\SI{1000}{m})=\SI{1.125}{m}$. The channel is discretized
into segments of length $\Delta
x=\SIlist[list-units=single]{1;5;10;50;100;200}{m}$. The Manning
roughness coefficient is set to $n=\SI{0.03}{m^{-1/3}s}$. The initial height
distribution is set to $y(x,t = 0) = \SI{0}{m}$. The numerical simulations
converge to steady-state after the time $t = \SI{4000}{s}$. 
The analytical and numerical solutions are compared in
 \ref{fig:delestre_3_3_1_9}. We obtain good agreement with a
maximum error of about \SI{\pm 1.8}{\percent} close to regions of
highest slope along the sinusoidal channel (see). For the lowest
resolution of $\Delta x=\SI{200}{m}$ absolute errors are slightly higher at
about \SI{6}{\percent}. The percentage RMSE drops to less than
\SI{0.7}{\percent} at the highest resolution and remains under
\SI{3}{\percent} for the lowest resolution at $\Delta x = \SI{200}{m}$.

\subsubsection{Steady free-surface flow with diffuse recharge}

To verify the implementation of the diffuse recharge we simulate flow in a
channel of length $L=\SI{1000}{m}$ with a specified constant inflow boundary
condition at the left boundary $x=\SI{0}{m}$ with $Q_f=\SI{1.0}{m^2 s^{-1}}$ and
a constant water depth at the right boundary with
$y=y_{eq}(\SI{1000}{m})=\SI{0.7483}{m}$. Here diffuse recharge refers to spatially distributed infiltration into the conduit system, typically originating from slow percolation through the epikarst or adjacent porous matrix. The channel segments are of length
$\Delta x=\SI{1}{m}$. The Manning roughness coefficient is set to
$n=\SI{0.033}{m^{-1/3}s}$. Diffuse recharge is applied to all nodes $0 \le x \le
L$ with $r_0 = \SI{0.001}{m s^{-1}}$. The steady-state water depth $y_0(x)$ is
given by the Gaussian-shaped function~\eqref{eq:analyical_delestre1}.

\begin{figure*}
\caption{Comparison of the numerical solution for different resolutions $\Delta x=$\SIlist[list-units=single]{1;5;10;25;50}{m} for the water depth at steady-state with the analytical solution Eq.~\eqref{eq:analyical_delestre1}. A steady diffuse recharge of $r_0 = \SI{0.001}{m s^{-1}}$ is applied along the whole domain. Notice that the channel base (lightgray) is only shown for a resolution of $\Delta x=\SI{1}{m}$. Errors are shown for all resolutions with lighter gray for higher values of $\Delta x$. RMSE values are computed based on the percentage errors.\label{fig:delestre_3_3_1_rainfall}}
\includegraphics[width=1\linewidth]{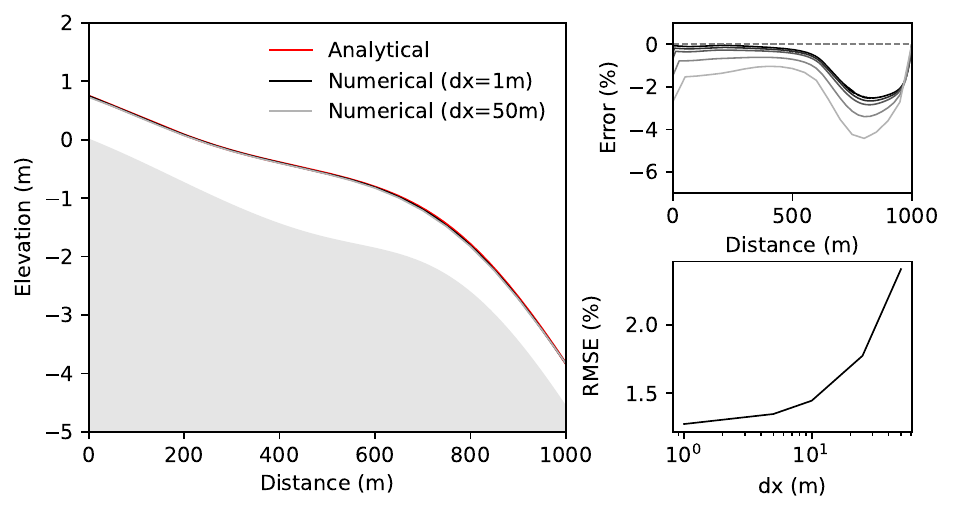}
\end{figure*}

The numerical model correctly recovers the depth profile along the channel with absolute
errors less than about \SI{\pm 4}{\percent} (see
 \ref{fig:delestre_3_3_1_rainfall}) at the highest resolution. Similar to
the previous examples the highest deviation can be observed close to the
steepest slopes of the system, except towards the right boundary where errors
are converging towards zero due to the applied constant head boundary. For the
lowest resolution of $\Delta x=\SI{50}{m}$ absolute errors are still below
\SI{6}{\percent}. Percentage RMSE errors drop below \SI{3.5}{\percent} with
increasing resolution and are at about \SI{4.6}{\percent} at the lowest
resolution of $\Delta x=\SI{50}{m}$.

\subsection{Steady-state flow in circular conduit under pressurized conditions}

 Here we consider steady-state flow through a horizontal conduit system
comprising ten conduit segments each with a length $\Delta x=\SI{100}{m}$, resulting
in a total length of $L=\SI{1000}{m}$. The conduits have a constant diameter of
$D=\SI{1}{m}$ and a base elevation of $z=\SI{0}{m}$. At the right boundary
($x=\SI{1000}{m}$) we maintain a constant water level $y=\SI{1.1}{m}$. At the
left boundary ($x=\SI{0}{m}$) we specify water depths in a range from
$y=\SI{1.15}{m}$ to $y=\SI{5}{m}$ to establish varying pressure gradients for
each equilibrium condition. Initial conditions are $y=\SI{0.9}{m}$ at the left
boundary and $y=\SI{0.8}{m}$ at the right boundary, i.e.,~flow is initially
non-pressurized. The initial discharge is $Q=\SI{0}{m^2 s^{-1}}$ and three
values for the Darcy--Weisbach roughness height are chosen as
$\epsilon=\SIlist[list-units=single]{0.001;0.01;0.1}{m}$. Equilibrium conditions
are computed at $t=\SI{4000}{s}$.

The numerical results are compared to the Darcy--Weisbach equation
\begin{equation}
\label{eq:DW_corrected}
    \frac{\Delta P}{L} = f_D\frac{\rho}{2D} \left(\frac{Q}{A_p}\right)^2
\end{equation}
where $A_p$ is the pressurized discharge area extended by the additional
contribution from the virtual slot (see Eq.~\eqref{eq:theta_discharge} and ~\eqref{eq:area_discharge}),
\begin{equation}
    A_p = \pi \frac{D^2}{4} + \frac{1}{N_c}\sum_{l=1}^N \left(\bar y_l - \Dc_l\right)
    W_0(\bar y_l),
\end{equation}
where $N_c=100$ is the total number of segments $l$.

\begin{figure*}
\caption{Comparison of steady-state discharge for three different roughness coefficients under pressurized flow conditions in a conduit of length $L=\SI{1000}{m}$ with $\Delta x=\SI{100}{m}$ (left) and corresponding errors (right). Circles represent the numerical solution obtained at an equilibrium at $t=\SI{2000}{s}$ and lines the analytical solution computed with Eq.~\eqref{eq:DW_corrected}.\label{fig:churchill_pressurized}}
\includegraphics[width=1\linewidth]{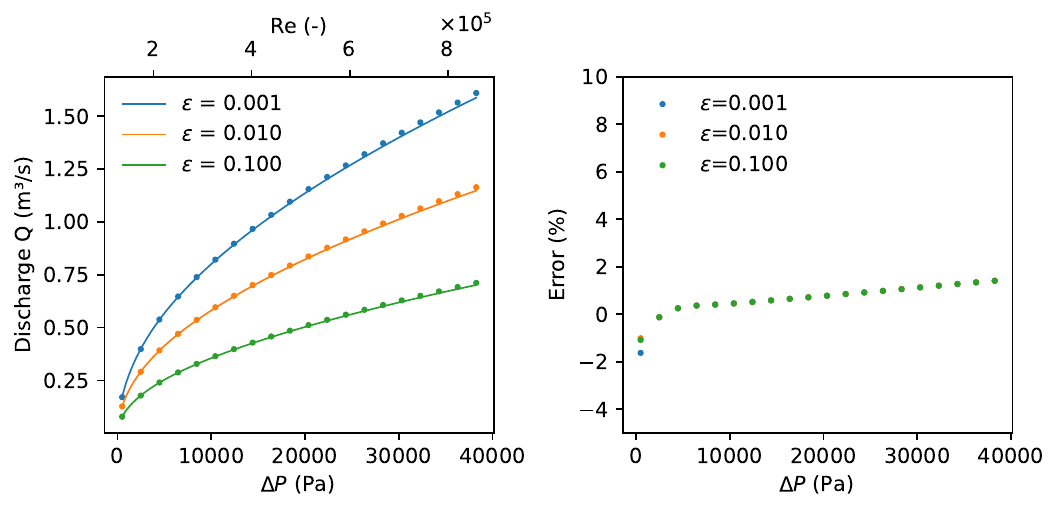}
\end{figure*}

Given a pressure differential $\Delta P$ along the distance $L$ based on
specified constant water depths at the left and right boundaries and the surface
roughness $\epsilon$ we iteratively solve Eq.~\eqref{eq:DW_corrected} to compute
the corresponding discharge $Q$.  \ref{fig:churchill_pressurized} shows the
numerical solution and the analytical solution for three roughness values. Good
agreement is found over the whole range of considered pressure drops and
Reynolds numbers. Errors exhibit a slight increase as Reynolds numbers rise but
remain below \SI{2}{\percent}. The stronger variation in errors for lower
pressure drops and Reynolds numbers can be explained by the chosen water depths
at the left boundary, which here result in most nodes of the conduit being below
or close to the critical height $\hat{y}_c$ so that the slot
width is determined according to Eq.~\eqref{eq:W_v} and becomes more sensitive to minor pressure variations.

\subsection{Transient free-surface flow with diffuse recharge}

Analytical solutions for the full dynamic wave model, which retains all terms of
the momentum equation Eq.~\eqref{eq:momentum}, are extremely challenging to
derive, even for simple geometries and boundary conditions. Therefore, various
simplifications have been proposed. For many flood routing applications in
natural channels, the inertial terms of the Saint-Venant equation are often
omitted, leading to the diffusive wave model~\citep{Hayami1951}. This single
parabolic equation captures only diffusive transport processes of the flow
dynamics. Wave velocity and diffusivity in this model are treated as
constants. A further simplification can be obtained by removing the water depth
gradient, which yields the so-called kinematic wave equation.

In order to validate the transient flow dynamics under consideration of all
terms of the momentum equation~\eqref{eq:momentum2} and allow for spatial and
temporal variation of wave velocity and diffusivity, we compare our code to the
laboratory experiment of~\citet{Delestre2013}. The experiment consists of an
inclined metal channel with a total length of $L=\SI{4}{m}$ and a width of
$b=\SI{0.12}{m}$. The slope of the channel is $\Sb=\SI{4.96}{\percent}$ and a
constant rainfall of $P=\SI{50.76}{mm h^{-1}}$ is applied along the channel
reach for times $t \in [\SI{5}{}, \SI{125}{s}]$ and within distance $x \in
[\SI{0}{}, \SI{3.95}{m}]$. In the numerical simulations the computation of the
discharge areas ($A= by$) and hydraulic radius are adapted to a rectangular
channel structure, that is,
\begin{equation}
    R = \frac{b y}{b + 2y}
\end{equation}
The channel is discretized with 100 segments of length $\Delta x=\SI{0.04}{m}$. The
base of the channel is dropping linearly from a height of $z = \SI{0.1984}{m}$
at $x = \SI{0.0}{m}$ to a height of $z = \SI{0.0}{m}$ at $x =
\SI{4.0}{m}$. Initial conditions are $y=\SI{0}{m}$ and $Q=\SI{0}{m^3 s^{-1}}$ at
all nodes. To simulate a outfall condition a constant water depth
$y=\SI{0.0}{m}$ is prescribed at the right boundary
($x=\SI{4.0}{m}$). Simulations are run with a time increment $dt=\SI{0.05}{s}$
to a maximum time of $\Delta t=\SI{250}{s}$. A constant precipitation $I_p =
\SI{1.44e-5}{m s^{-1}}$ is applied at times $\SI{5}{s} < t < \SI{125}{s}$,
resulting in a volumetric inflow of $Q_r =\SI{1.44e-5}{m s^{-1}} \times
(\SI{0.12}{m} \times \SI{0.04}{m}) = \SI{6.933e-8}{m^3 s^{-1}}$ at all channel
segments. Note that, in line with~\citet{Delestre2010} a corrected precipitation
of $I_\mathrm{corr}=\SI{52}{mm h^{-1}}$ is assumed in the simulations.

\begin{figure*}
\caption{Comparison of the numerical solution for the discharge and cumulative discharge with the experimental data of~\citet{Delestre2010} under consideration of a constant diffuse recharge of $R_c = \SI{52}{mm h^{-1}}$ for
$\SI{5}{s} < t < \SI{125}{s}$ along the channel between $x=\SI{0}{m}$ and $x=\SI{3.95}{m}$. The channel is $L=\SI{4}{m}$ long with a slope of $\Sb = \SI{4.96}{\percent}$, $\Delta x=\SI{1}{m}$ and $\Delta t=\SI{0.05}{s}$. A constant water depth of $y=\SI{0.0}{m}$ is prescribed at
the right boundary $x=\SI{4.0}{m}$. The Manning coefficient is set to $n=\SI{0.013}{m^{-1/3}s}$ and initial conditions are $y=\SI{0}{m}$ and $Q=\SI{0}{m^2 s^{-1}}$.\label{fig:delestre_6_1_rainfall}}
\includegraphics[width=1\linewidth]{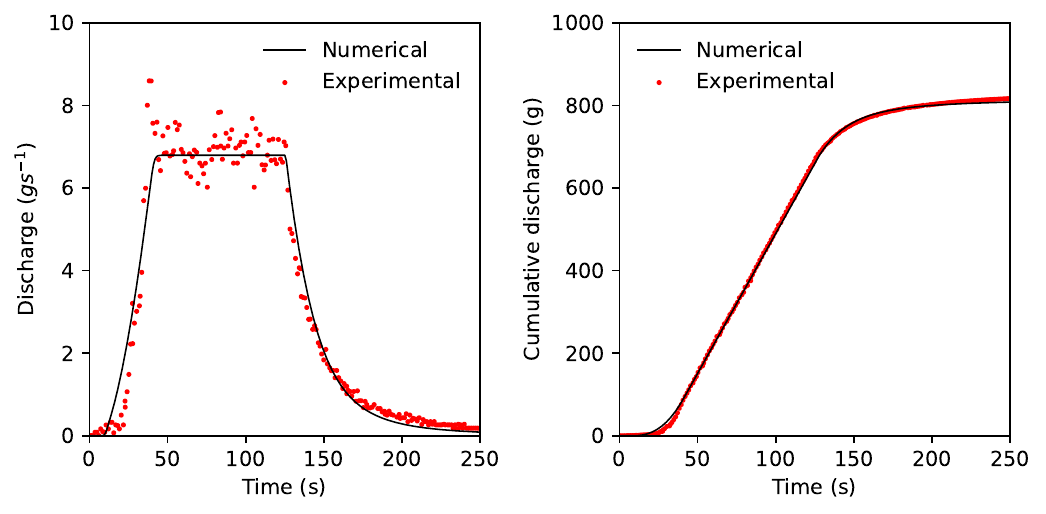}
\end{figure*}

Similar to~\citet{Delestre2010} we find that a Manning roughness coefficient of $n=\SI{0.013}{m^{-1/3}s}$ reproduces the experimental data with satisfactory accuracy (see  \ref{fig:delestre_6_1_rainfall}). The initial peaks in
discharge at about $\Delta t=\SI{40}{s}$ and fluctuations along the period of constant discharge are most likely a result of preferential flow formation, i.e.~instabilities along the channel width. Due to the chosen 1D modeling
approach they cannot be recovered in the simulations, which was also noted by~\citet{Ersoy2020}. The cumulative discharge mass also shows good agreement with the experimental data and closely aligns with the results
obtained by~\citet{Delestre2010} and~\citet{Ersoy2020}.

\section{Demonstration example: Ox Bel Ha cave system, Quintana Roo, Mexico}
\label{sec:demonstration}
Here, we demonstrate the simulation of flow in the Ox Bel Ha system, one of the longest known submerged cave networks worldwide, located in Quintana Roo, Mexico. As of 2025, the officially reported length of the explored
portion of the Ox Bel Ha cave system is approximately \SI{524.1}{km}. The average depth of the system is around \SI{16}{m}, with a maximum recorded depth of \SI{57.3}{m}. It should be noted that the dataset used in this
study represents an earlier version with a total mapped length of \SI{435.8}{km}. The cave system is connected to the surface via 152 cenotes (sinkholes) and has several discharge outlets towards the Caribbean
coast~\citep{Oxbelha2023}.

The modeled network was derived from available survey data and comprises 10098 conduits with lengths ranging from \SI{0.6}{m} to \SI{111.52}{m}. It represents only a portion of the physically explored cave system and serves as a simplified approximation for simulating flow dynamics. Due to the limited availability of calibrated field data, we do not attempt a comparison with observations here. This demonstration highlights the scalability of the code and its capability to simulate fully transient flow in large, real-world karst networks. For the sake of simplicity and due to the absence of detailed data, we assume an average diameter of all conduits of \SI{1}{m} and a uniform roughness height of $\epsilon = \SI{0.03}{m}$. For the transient simulations, recharge is applied at designated injection nodes that represent cenotes or sinkholes, which are distributed at regular intervals along the modeled network based on nodal indices. This uniform distribution is a simplification used for demonstration purposes. Nodes close to the Caribbean coast are set as constant pressure outlets. Their water depths are adjusted to a constant value in order to maintain a constant hydraulic head of $H=\SI{2}{m}$, which reflects the constant elevation of the ocean during the injection scenario.

As starting conditions, the conduit network is brought to steady-state conditions by applying a total recharge of \SI{0.02}{m^3 s^{-1}}, which is uniformly distributed across all injection nodes. This approximately corresponds to an average annual recharge of \SI{180}{mm/a} given a catchment area of about \SI{36}{km}. We then simulate a heavy storm event with an intensity of \SI{40}{mm/h}. We model the input signal at each sinkhole over a time period of \SI{2}{h} as a linear ramp to \SI{0.05}{m^3s^{-1}} at \SI{1}{h} before decreasing again until \SI{2}{h} (see  \ref{fig:oxbelha_numericalresults}, upper left). For an average sinkhole diameter of \SI{75}{m} this corresponds to an injection rate of about \SI{0.05}{m^3s^{-1}}. Due to the shallow depth of the conduit system and the rapid transfer of recharge through the sinkholes, we assume an instantaneous injection at the conduit level.

 The simulation employs a constant time increment $\Delta t=\SI{0.025}{s}$, a relaxation factor $\omega=0.6$ and a Picard tolerance of $\delta = \SI{10e-5}{m}$. It was run single-threaded on an Apple M3 Pro SoC (11-core: 5P + 6E), whose P-cores reach up to 4.06~GHz (E-cores ~2.7~GHz), with \SI{32}{GB} memory ($\sim$150~GB/s) and completed in \SI{89}{min} with a total of 86400 time steps. \SI{87.4}{\percent} of the time steps required two iterations and \SI{12.6}{\percent} required 3 iterations to converge, with no step hitting the iteration cap and no convergence failures.

\begin{figure*}
\caption{Flow results of the Ox Bel Ha cave network recharge scenario. Conduit colors represent flow rates and vertical lines the nodal water depths. The left figure shows the system close to steady-state conditions when the injection pulse starts ramping up (see  \ref{fig:oxbelha_numericalresults}, upper left). Green spheres indicate the sinkhole locations, red spheres the outlets. The right figure shows the system at maximum outlet discharge, which occurs at about \SI{50}{min} after the maximum peak of the recharge signal.\label{fig:oxbelha_paraview}}
\includegraphics[width=1\linewidth]{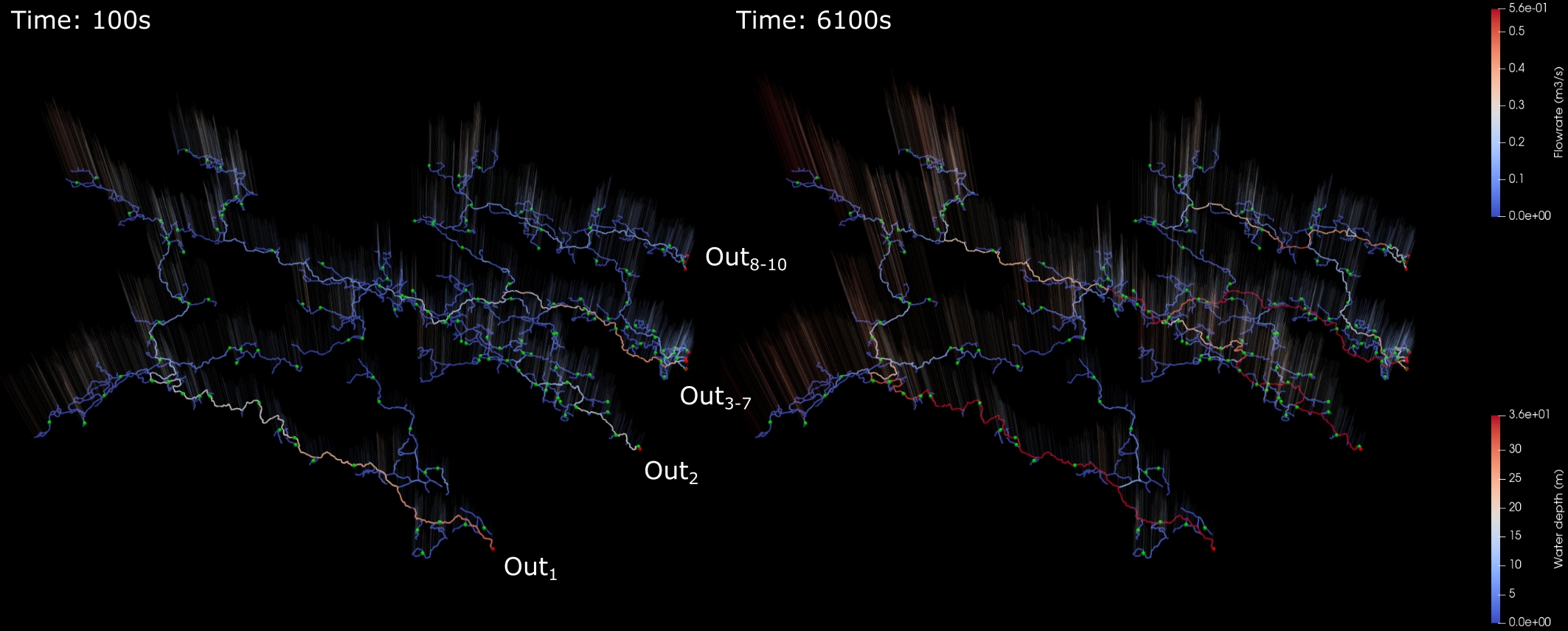}
\end{figure*}

\begin{figure*}
\caption{(Upper left) Discharge at various outflow nodes along the eastern coastline (see  \ref{fig:oxbelha_paraview}), where outflows for close nodes have been combined. The recharge signal (right y-axis) peaks at \SI{1}{h} with a maximum of \SI{7.59}{m^3s^{-1}}, i.e.,~the total recharge applied to all 152 sinkholes. (Lower left) Fraction of conduits that are in a pressurized state. (Upper right) Fraction of conduits that exhibit turbulent flow. (Lower right) Volume ratio of the main pathways identified based on the percentile of flow rates, indicating a high flow concentration in the network.\label{fig:oxbelha_numericalresults}}
\includegraphics[width=1\linewidth]{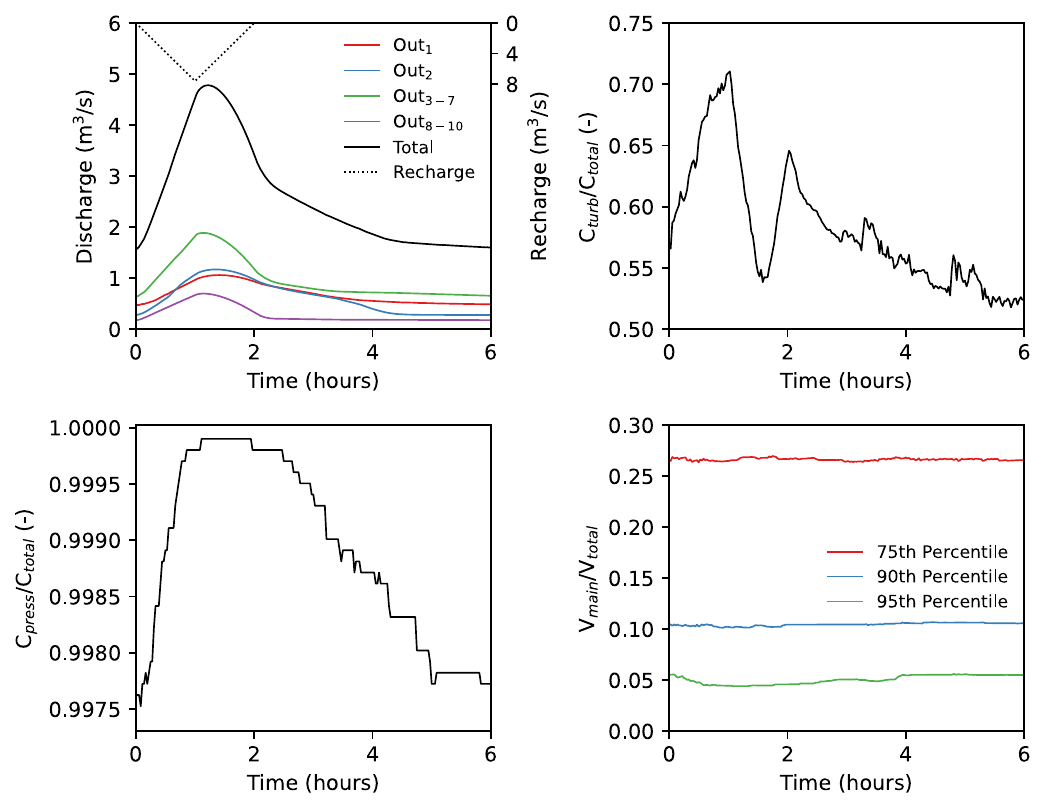}
\end{figure*}

 \ref{fig:oxbelha_paraview} shows the Ox Bel Ha cave system shortly after the onset of the recharge event at $t=\SI{100}{s}$ (left figure) and at time $t=\SI{6100}{s}$ when the peak discharge at the spring outlets is registered (right figure). Flow rate distributions indicate two main pathways connected to outlets 1 and 3--7 close to steady-state conditions, while outlets 2 and 8--10 are activated more efficiently during peak flow conditions. At outlets 8--10 this is most likely caused by the marginal position of the branch and connectivity to the main branches in southwestern direction, which is oriented against the main flow gradient towards the southwest. Outlet 2 on the other hand is seemingly well connected to the upstream network in northwestern direction, similar to outlets 3--7. However, the majority of conduits, up to the position where the branch merges with the main branch connected to outlet 3--7, are about \SI{10}{m} deeper. Therefore, the branch system connected to outlet 2 forms a trough that can accommodate larger water volumes during the recharge event, and hence is activated with a slight delay.

These processes can also be observed in  \ref{fig:oxbelha_numericalresults} which shows the discharge, the fraction of pressurized conduits, the fraction of turbulent conduits, and the ratio of main branches to total branches. In terms of outflow volume, the branch connected to outlet 3--7 receives the highest amount of discharge due to its central location and hence highest contributing upstream conduit volume, while outlets 8--10 receive the lowest amount of recharge from the upstream network. The activation of the previously discussed trough area connected to outlet 2 can be observed in the graph showing the fraction of turbulent conduits (upper right). These range from about \SI{50}{\%} under steady-state conditions to a maximum of about \SI{70}{\%} at the maximum injection rate after \SI{1}{h}. A second maximum after about \SI{2}{h} can be attributed to the filling of the trough area connected to outlet 2 and, furthermore, the delayed filling of the northeastern branches.

The whole network is nearly fully pressurized over the whole simulation period. Only about \SI{0.25}{\%} of conduits located close to the merging point of main branches connected to outlet 2 and outlet 3--7 are in a free-surface flow mode at steady-state (lower left). Here a local maximum of the conduit base heights is observed while conduit elevations drop towards the northeastern coastline and the northwestern parts. In order to quantify the flow focusing properties of the network, we compute the time-dependent ratio of main branch volume to total conduit volume based on the percentile of flow rates (lower right graph). The network exhibits a very strong degree of flow focusing. The top \SI{5}{\%} of conduits with highest flow rates make up partially less than \SI{5}{\%} of the total conduit volume, and hence form dominant pathways.

\section{Conclusion and future directions}
\label{sec:conclusion}
 In this paper, we describe the development of \texttt{openKARST}, a new Python-based code for the simulation of highly dynamic flows in complex karst conduit networks via the dynamic wave equation. \texttt{openKARST} is written in an efficient vectorized form and provides all basic functionalities for the systematic investigation of the impact of network geometry and heterogeneity on the flow dynamics in karst networks. The code simulates free-surface flow and dynamic transitions to fully pressurized conditions both under laminar and turbulent conditions governed by the Darcy--Weisbach and Manning equations using a continuous Churchill friction factor formulation.

The \texttt{openKARST} codebase includes a variety of ready-to-run examples that cover synthetic setups, real cave data, and comparisons with analytical solutions. The examples and manual demonstrate how to configure simulations via intuitive dictionary-based settings, apply time-dependent or spatially distributed boundary conditions, and visualize results using built-in animation tools or export functionalities (e.g.,~VTK, PyVista, Plotly/Dash). A more detailed description of a typical user interaction, including pre- and post-processing and a minimal working example, is provided in  \ref{app:user}. The code has been carefully verified and validated for steady-state and transient flow dynamics. Finally, we have used the code to solve for flow in the Ox Bel Ha cave network system and assess the influence of network topology and geometry on transient recharge and integrated discharge dynamics.

While current applications demonstrate the capabilities of the model in synthetic and survey-based networks, future comparisons with monitored discharge and pressure data will be critical to further demonstrate the applicability to other real-world systems. \texttt{openKARST} and established tools such as MODFLOW-CFP, MODBRANCH, and ModBraC all simulate flow in karst or related hydrogeological systems, but they were developed with different modeling goals and temporal scales in mind. MODFLOW-CFP is well suited for long-term simulations at catchment scales where matrix--conduit exchange dominates and assumes steady or quasi-steady conduit flow. MODBRANCH and ModBraC extend MODFLOW to solve the Saint-Venant equations, primarily to address stream--aquifer or matrix--conduit interactions within the MODFLOW framework. In contrast, \texttt{openKARST} provides a stand-alone Python implementation of the fully transient Saint-Venant equations, making it particularly suited to capture rapid flow events such as flood waves, recharge pulses, and dynamic pressurization in complex conduit networks.

Future directions include the implementation of solute transport on different levels of complexity via the advection--dispersion equation, random walk particle tracking, and time-domain random
walks~\citep{Noetinger2016}, improved geometry descriptors based on cave surveys to replace circular geometries~\citep{Collon2017}, and coupling of flow and transport to the porous matrix systems via the source--sink terms
in the continuity and momentum equations. In this context, \texttt{openKARST} is highly suited for future coupling of conduit flow with fractured porous media flow models. Such integration will allow modeling of exchange
processes between flow in conduits and the surrounding porous matrix domain, as for example explored in recent works on discrete fracture--cave systems~\citep{Zhang2022}. Fully-coupled model approaches enable the
investigation of permeability evolution, flow channeling, and recharge partitioning across compartments, and allow to bridge the gap between network-scale flow modeling and reservoir-scale characterization of fractured
karst systems~\citep{Berre2019,Jourde2023}.

The authors acknowledge funding by the European Union (ERC, KARST, 101071836). Views and
opinions expressed are, however, those of the authors only and do not

necessarily reflect those of the European Union or the European Research Council
Executive Agency. Neither the European Union nor the granting authority can be
held responsible for them. Regarding the Ox Bel Ha network data, we thank James G. Coke for preparing and sharing the survey data with us, as well as the GEO (Grupo de Exploraci\'on Ox Bel Ha), the MCEP (Mexico Cave Exploration Project), the QRSS (Quintana Roo Speleological Survey), the CINDAQ (El Centro Investigador del Sistema Acu\'\i fero de Quintana Roo), and all the cave surveyors who have explored and mapped the cave over the years.

\subsection*{Code availability section}

Name of the code/library: openKARST

Contact: Jannes Kordilla, jannes.kordilla@idaea.csic.es

Hardware requirements: 64-bit

Program language: Python 3

Software required: Windows, UNIX/Linux, macOS

Program size: 4.6 MB (including example data) 

The source code and analytical solutions are available at \url{https://doi.org/10.5281/zenodo.16794329}[Zenodo (DOI: 10.5281/zenodo.16794329)]
and on GitHub at \url{https://github.com/ERC-Karst/openkarst}[github.com/ERC-Karst/openkarst].

License type: GPL

\appendix

\section{Adaptive time increment}
\label{app:time_step}

 The time increment is computed according to an adaptive scheme based on the value of the local Froude number $\Fr_l$~\citep{SWMM2017}. To this end, we define the time increment
\begin{equation}
    \Delta t_{l} \le \Cr \left(\frac{\Fr_l}{1+\Fr_l}\right)\frac{\Dc_{l}}{\vert \bar v_l\vert },
\end{equation}
where $\Dc_{l}$ is the hydraulic diameter of conduit $l$ and $\Cr$ the Courant
number, which is set here equal to $1$ for free-surface and to $1/2$ for
pressurized flow. Highly dynamic flows often require stricter constraints,
typically setting (free-surface flow) $\Cr < 1$ and (pressurized flow) $\Cr < 1/2$
to ensure numerical stability. For steadier and less erratic flow dynamics, it
may be feasible to relax the constraints and set $\Cr \geq 1$ without losing
accuracy in the simulation. Furthermore, we define
time increment $\Delta t_{i}$ based on the rate of change in water depth
relative to the maximum conduit diameter $D^m_{i} = \max_l(\Dc_{i,l})$ connected to
the node
\begin{equation}
    \Delta t_i \leq \Delta t_{k} \frac{D^m_{i}}{y_i^{k+1} - y_i^k},
\end{equation}
where $\Delta t_k$ is the current time increment. The time increment $\Delta t_{k+1}$
at the time step $k+1$ is then determined as the minimum value of all $\Delta
t_l$ and $\Delta t_i$,
\begin{equation}
    \Delta t_{k+1} = \min_{l,i}(\Delta t_l,\Delta t_i). 
\end{equation}
 
\section{Closure relations}
\label{app:closures}

 In the following, we provide the closure relations needed for the numerical solutions of Eqs.~\eqref{eq:nodal_volume_discrete} and ~\eqref{eq:momentum_discrete}. 
 
\subsection{Computation of free-surface and discharge areas}
\label{sec:surfaces}

\begin{figure}
\centering
\caption{Computation of hydraulic parameters for a pressurized
      system including a Preissmann slot (top) and free-surface system
      (bottom) of two conduits connected at node $y_i$. The Preissmann
      slot is not to scale. Note that other combinations of
      pressurized and free-surface conditions may exist and more than
      one conduit may be connected to a node. Switching to pressurized
      flow computation using the Darcy--Weisbach equation is based on
      the pressurization state at $\bar{y}$ only.\label{fig:conduit_discretization}}
\includegraphics[width=0.5\linewidth]{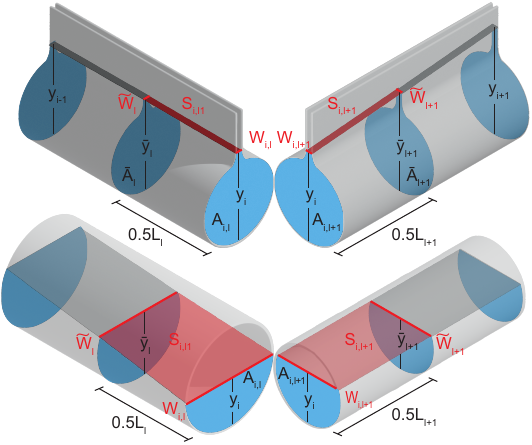}
\end{figure}
 The computation of nodal free-surface areas to update water depths via
Eq.~\eqref{eq:nodal_volume_discrete}, conduit discharge areas and hydraulic
radii required in Eq.~\eqref{eq:momentum_discrete} represents one of the most
computationally demanding tasks within the code. The hydraulic parameters depend on the specific geometry of the conduits. For any cross-sectional shape the
surface area $A$, surface width $W$ and the perimeter $P$ are a
function of the water depth $y$ (see  \ref{fig:conduit_discretization}). The
functional relations may strongly vary for circular, rectangular or trapezoidal
geometries. For more complex geometries there may not be closed form
analytical expressions. In these case, $A$, $W$ and $R$ are determined
numerically and their dependence on $y$ is tabulated.

\begin{figure*}
\centering
\caption{(Left) Evolution of the slot width under pressurized
      conditions for a conduit with diameter $D=\SI{1}{m}$. A constant
      slot width of $W_0 = 0.01D$ is reached at
      $y_c=1.78 \Dc$. (Right) Evolution of the discharge area in a
      conduit with $\Dc=\SI{1}{m}$. Beyond a normalized depth of
      $y = \Dc$, the additional contribution is due to the
      Preissmann slot.\label{fig:slotwidth}}
\includegraphics[width=1.0\linewidth]{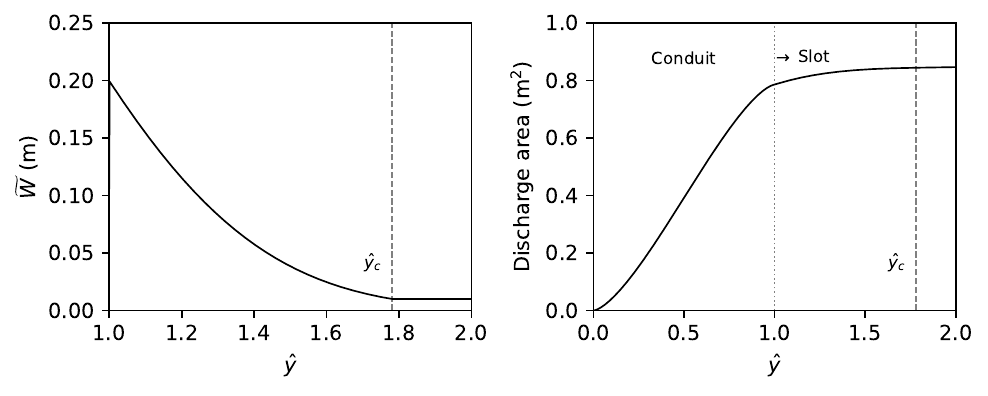}
\end{figure*}

In the current code version we implement circular and rectangular conduits with
constant diameter $\Dc$ or width $b$ and length $L$.
For the rectangular conduits, the width of the free-surface, discharge area,
wetted perimeter and hydraulic radius are given by
\begin{equation}
W(y) = b, \qquad A(y) = b y, \qquad P(y) = b + 2y, \qquad R(y) = \frac{by}{b + 2y}.   
\end{equation}

For circular conduits, the free-surface width $W$ is expressed in terms of the water depth $y$ as
\begin{equation}
W(y) = 
\begin{cases} 
W_0 & \text{if } y \ge \Dc \\
2 \sqrt{\Dc y - y^2} & \text{if } y < \Dc 
,\end{cases}\end{equation}
where $W_0$ is the virtual surface width, or Preissmann slot
\citep{Preissmann1961}, required to deal with pressurized flow conditions as
discussed in the next section. The discharge area $A$ is determined in terms of the water depth as
\begin{equation}
\label{eq:area_discharge}
A(y) = 
\begin{cases} 
\pi \Rc^2 + (y - \Dc)  W_0 & \text{if } y \ge \Dc \\
\Rc^2\frac{\theta(y) - \sin{\theta(y)}}{2} & \text{if } y < \Dc 
,\end{cases}\end{equation}
where $\Rc$ [L] is the radius of the conduit and
\begin{equation}
\label{eq:theta_discharge}
\theta(y) = 
\begin{cases} 
2 \pi & \text{if } y \ge \Dc \\
2 \arccos\left(\frac{\Rc - y}{\Rc}\right) & \text{if } y < \Dc 
.\end{cases}\end{equation}
The wetted perimeter $P$ and hydraulic radius are
\begin{equation}
P(y) = \Rc \theta(y), \qquad R(y) = \frac{A(y)}{P(y)}. 
\end{equation}

\begin{figure}
\centering
\caption{Friction factor f for a range of Reynolds numbers and three values of the relative roughness $\epsilon/D$. Solid lines represent the continuous Churchill equation, while dashed lines the Colebrook--White equation. The shaded area marks the transition region between laminar and turbulent conditions ($2300 \ge Re \ge 4000$).\label{fig:churchill_analytical}}
\includegraphics[width=0.5\linewidth]{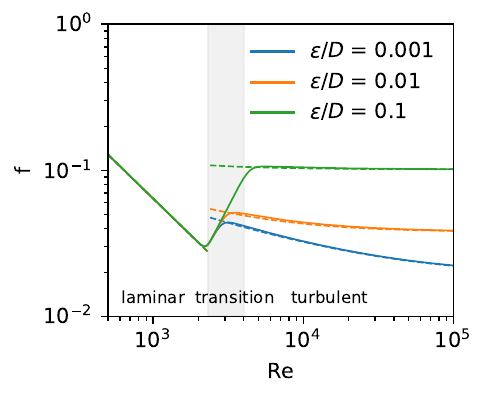}
\end{figure}
  
\subsection{Preissmann slot}
\label{subsec:pressurized}
 When nodes or conduit centers become pressurized, that is, if the water depth
$y_i$ at a node exceeds the maximum diameter of all the connected conduits, or
when $\bar{y}_l$ at a conduit center exceeds the conduit diameter, the
computation of the free-surface areas $S$ and thus also the discharge area
$A$ must be modified. A conduit is considered pressurized if $\bar y_l \ge
\Dc_{l}$, a node is considered pressurized if the nodal water depth is larger than or
equal to the maximum diameter of the conduits that are connected to it, $y_i \ge
\max_l(D_{i,l})$. As the pressurization state of a conduit is determined based on the average water
depth $\bar{y}_l$, it can happen that the conduit is pressurized at one end (if
$y_i > \Dc_{i,l}$), while it is considered in free-surface flow mode because $\bar y_l
< \Dc_l$. The flow mode of the conduit is important for the switching between
two different friction slopes, as detailed in the next section. 

If a node and the connected conduits pressurize, the nodal free-surface area
$S_{i}$ defined by Eq.~\eqref{eq:nodal_area} becomes zero. Thus, it is not
possible to update the nodal water depth $y_i$ using
Eq.~\eqref{eq:nodal_volume_discrete}. In order to circumvent this problem, a
virtual water depth and hence an extended free-surface area is computed
by assigning a finite value to the free-surface width $W = W_0$ when $y
\ge \Dc$, which is the width of the so-called Preissmann
slot~\citep{Cunge1964}. Adding this virtual storage is equivalent to
considering the water as slightly compressible.
 
For $y > \Dc$ but below a critical threshold, the virtual slot widths $W_0$ is
determined following~\citet{Sjoberg1982} and~\citet{SWMM2017}. When $y$ exceeds
a critical value of $y_c = 1.78 \Dc$, the slot width
is set equal to 1\% of the conduit diameter $\Dc$ (see
 \ref{fig:slotwidth}, left). That is,
\begin{equation}
\label{eq:W_v}
W_0(y,\Rc) = 
\begin{cases} 
 0.5423 \exp\left(-(y/\Dc)^{2.4}\right) d & \text{for } \Dc \leq y \leq y_c\\
0.01 D & \text{for } y > y_c
.\end{cases}\end{equation}
If both node and connected conduits are pressurized, the nodal free-surface
area $S_i$ is non-zero because it is computed
in terms of the virtual slot widths $W_0(y_i,\Rc_l)$ and $W_0(\bar
y_{i,l},\Rc_{i,l})$. The respective discharge areas for pressurized flow are
extended by the additional area due to the slot above the conduit ceiling as given in
Eq.~\eqref{eq:area_discharge}, see also  \ref{fig:slotwidth}. The hydraulic
perimeter and hydraulic radius are not recalculated for the Preissmann slot
because its impact is negligible.

\subsection{Friction slope}
\label{sec:friction}
 The friction slope can be quantified by the Manning or Darcy--Weisbach equations. The Manning formula for the friction slope
reads as
\begin{equation}
\label{eq:manning_sf}
    \Sf = \frac{h_f}{L} = n^2 \frac{Q \vert v \vert}{AR^{4/3}},
\end{equation}
where $h_f$ [$L$] is the head loss along distance $L$ [$L$], $n$ is the Manning
friction coefficient [$T L^{-1/3}$] and $R$ is the hydraulic radius [$L$]. The
Darcy--Weisbach equation results in the following friction slope,
\begin{equation}
\label{eq:dw_frictionslope}
    \Sf = \frac{f_D v^2}{8gR}.
\end{equation}
where $f_D$ is the dimensionless friction coefficient.  Under laminar conditions ($Re < 2300$) the friction factor is
\begin{equation}
    f_D = \frac{64}{Re},
\end{equation}
where the Reynolds number is defined as
\begin{equation}
    \R = \frac{\rho \vert \bar{v} \vert D}{\mu}
\end{equation}
 with $\rho=\SI{1000}{m^3 s^{-1}}$ the water density and $\mu =
\SI{0.001}{kg m^{-1}s^{-1}}$ the dynamic viscosity.
For turbulent conditions, various models such as the Colebrook--White equation
\citep{Colebrook1937} or Swamee--Jain formulation~\citep{Swamee1976} have been
proposed to solve for the friction factor, i.e.,~when $Re > 2300$. As these
formulations implicitly depend on $f$ they are often solved iteratively. Here we
employ an alternative formulation after~\citet{Churchill1977}, which allows a
continuous explicit calculation of the friction factor under both laminar and
turbulent conditions (see  \ref{fig:churchill_analytical}):
\begin{align}
f_D &= 8\left[ \left(\frac{8}{Re}\right)^{12} + \frac{1}{\Omega_1 + \Omega_2}^{\frac{3}{2}} \right]^{\frac{1}{12}}\\
\Omega_1 &= \left[-2.457 \ln\left({\left(\frac{7}{Re}\right)^{0.9} + 0.27\frac{\epsilon}{d}}\right)\right]^{16}\\
\Omega_2 &= \left(\frac{37530}{Re}\right)^{16},
\end{align}
where $\epsilon$ [$L$] is the effective conduit roughness height and
$\epsilon/D$ is referred to as the relative roughness [$-$].

Under free-surface flow conditions, the friction slope in
Eq.~\eqref{eq:momentum_discrete} is set equal to the Manning
formula~\eqref{eq:manning_sf} and under pressurized conditions, it is set equal
to the Darcy--Weisbach formula \eqref{eq:dw_frictionslope}. 
In order to allow for a continuous computation during transitions from free-surface
to pressurized conditions with a single descriptor for the roughness properties
we define a Manning coefficient that is consistent with the Darcy--Weisbach
friction factor. Thus, we equate Eqs.~\eqref{eq:dw_frictionslope} and ~\eqref{eq:manning_sf}
for the same velocity $v$ in order to derive
\begin{equation}
\label{eq:n}
    n = \sqrt{\frac{f_D R^{\frac{1}{3}}}{8g}}. 
\end{equation}
Note that this equation implies that the Manning coefficient depends on $\R$
through the dependence of the Darcy friction factor $f_D$ on $\R$. Thus, we define a
unique Manning coefficient by setting $f_D = f_D(\R = \infty)$ in
Eq.~\eqref{eq:n}. 
 
\section{Derivation of Eq.~\eqref{eq:momentum2}}\label{app:derivation}\label{app:numerics}

 In order to derive Eq.~\eqref{eq:momentum2}, we combine the continuity equation~\eqref{eq:continuity} and ~\eqref{eq:momentum} as follows. First expand the inertia term in Eq.~\eqref{eq:momentum} as
\begin{equation}
    \frac{\partial}{\partial x}\left(\frac{Q^2}{A}\right) = 2 \frac{Q}{A} \frac{\partial Q}{\partial
  x} - \frac{Q^2}{A^2} \frac{\partial A}{\partial x}. 
\end{equation}
Using the definition $Q = vA$, we can write this expression as
\begin{equation}
    \frac{\partial}{\partial x}\left(\frac{Q^2}{A}\right) = 2 v \frac{\partial Q}{\partial
  x} - v^2 \frac{\partial A}{\partial x}. 
\end{equation}
Per the continuity equation~\eqref{eq:continuity}, the spatial derivative of $Q$ is
\begin{equation}
\frac{\partial Q}{\partial x} = - \frac{\partial A}{\partial t} + q. 
\end{equation}
Thus, we obtain for the inertia term
\begin{equation}
    \frac{\partial}{\partial x}\left(\frac{Q^2}{A}\right) = - 2 v \left(\frac{\partial A}{\partial t} - q \right) - v^2 \frac{\partial A}{\partial x}. 
\end{equation}
Inserting this expression for the inertia term in Eq.~\eqref{eq:momentum} gives Eq.~\eqref{eq:momentum2}.

\section{User interaction, pre- and post-processing}
\label{app:user}

\texttt{openKARST} is designed to be used in scripts or Jupyter notebooks. A typical workflow for running simulations consists of five steps: (1) Construction of a conduit network (imported or synthetic), (2) providing fundamental simulation parameters such as physical properties (e.g.,~water density, viscosity, gravity), solver settings (e.g.,~relaxation factor, Picard tolerances, Courant-limited time-step criteria), and desired output frequency and variables, (3) setting of initial and boundary conditions, (4) running the solver, and lastly (5) storing, exporting or visualizing results. In the following, we illustrate the main pre- and post-processing options and include a minimal usage example.

\subsection{Pre-processing}

Karst conduit networks can be imported from surveyed cave data or generated (e.g.~standard cubic lattices or regular grids). All functionalities to interact with the geometry object available in OpenPNM~\citep{Gostick2016}
can be employed in \texttt{openKARST}. This includes assigning properties (conduit diameters, roughness), determining node locations (e.g.~for boundary conditions) and validating or correcting connectivity. Boundary
conditions can be defined at nodes via methods associated with the flow simulation object, which support various constant and time-dependent input options for inflow and water depth (e.g.user-defined time series, ramps or
step functions).

\subsection{Minimal usage example}

Below we provide a minimal working example, which demonstrates the setup of a short linear network with a single inflow and one water-depth boundary condition, export of results to VTK, and finally visualization via the browser-based 3D \texttt{openKARST viewer}.

\begin{lstlisting}[style=terminal, caption={Minimal example}]
import openpnm as op
import numpy as np
from openkarst.network_generation import compute_conduit_lengths
from openkarst.models import FlowSimulation
from openkarst.io.vtk_data_exporter import VtkDataExporter
from openkarst.visualization.openkarst_viewer import launch_openkarst_viewer

# Create geometry object (200 nodes, 1 m spacing)
geo = op.network.Cubic(shape=[200, 1, 1], spacing=1.0)
geo = compute_conduit_lengths(geo)
geo['throat.diameters'] = 1.0
geo['throat.epsilon']   = 0.03

# Settings (some defaults omitted for brevity)
physical = {'water_density':1000, 'gravity':9.81, 'dynamic_viscosity':1e-3}
solver   = {'relaxation_factor':0.6, 'max_iterations':20}
sim      = {'courant':0.8, 'adaptive_timesteps':True,'steady_state':True,'t_init':0.1}
outputs  = {'output_interval':10.0, 'time':True, 'flowrates':True, 'water_depths':True}

# Model, ICs, BCs
model = FlowSimulation(geo, physical, solver, sim)
model.set_initial_conditions(np.zeros(geo.Nt), np.full(geo.Np, 0.01))
model.set_inflow_BC(nodes=[0],    values=0.10, inflow_type='volumetric')
model.set_waterdepth_BC(nodes=[199], values=0.01)

# --- Observation nodes (record 'water_depth' and 'inflow' every 1 s)
model.set_observation_points(nodes=[0, 100, 199],
                             variables=['water_depth', 'inflow'],
                             interval=1.0)

# Run and export to VTK
res = model.run_simulation(desired_outputs=outputs)
VtkDataExporter('vtk_output').export(geo, res['flowrates'], res['water_depths'], res['time'])

# Collect observations as a pandas.DataFrame
obs_df = model.observation_recorder.to_dataframe()

# Interactive 3D viewer
# Opens a browser window for interactive playback
launch_openkarst_viewer(results=res, geometry=geo, observations=obs_df)
\end{lstlisting}

\subsection{Post-processing and visualization}

In addition to standard visualization options via commonly applied Python libraries, results can be exported to VTK for ParaView, animated with PyVista, or interactively visualized in the included Dash/Plotly 3D viewer. Observation points can be set to record time series of node variables with the desired frequency, and also be visualized via the 3D viewer.

\subsection{Reproducible examples and manual}

The code provided in the Zenodo repository comes with a user manual featuring a set of detailed examples covering (1) flow in synthetic karst conduit networks, (2) simulation of flow in imported real cave networks, and (3) examples of analytical benchmarks.

\printcredits

\bibliographystyle{cas-model2-names}

\bibliography{library}

\nomenclature[A]{$A$}{Wetted discharge area [L$^2$]}
\nomenclature[B]{$A_p$}{Pressurized discharge area [L$^2$]}
\nomenclature[C]{$\Ac$}{Conduit cross-sectional area [L$^2$]}
\nomenclature[D]{$\alpha$}{Upstream weighting factor [--]}
\nomenclature[E]{$b$}{Channel width [L]}
\nomenclature[F]{$Cr$}{Courant number [--]}
\nomenclature[G]{$\Dc$}{Conduit diameter [L]}
\nomenclature[H]{$d_i$}{Degree of node $i$ [--]}
\nomenclature[I]{$\delta$}{Picard convergence tolerance [--]}
\nomenclature[J]{$\Delta P$}{Pressure difference [M L$^{-1}$ T$^{-2}$]}
\nomenclature[K]{$\Delta t$}{Time increment [T]}
\nomenclature[L]{$\Delta x$}{Spatial resolution [L]}
\nomenclature[M]{$\epsilon$}{Conduit roughness height [L]}
\nomenclature[N]{$f_D$}{Darcy–Weisbach friction coefficient [--]}
\nomenclature[O]{$Fr$}{Froude number [--]}
\nomenclature[P]{$g$}{Gravitational acceleration [L T$^{-2}$]}
\nomenclature[Q]{$H$}{Hydraulic head [L]}
\nomenclature[R]{$h_f$}{Head loss [L]}
\nomenclature[S]{$I_p$}{Precipitation rate [L T$^{-1}$]}
\nomenclature[T]{$i$}{Node index [--]}
\nomenclature[U]{$j$}{Picard iteration index [--]}
\nomenclature[V]{$k$}{Time step index [--]}
\nomenclature[W]{$l$}{Conduit index [--]}
\nomenclature[X]{$L$}{Channel or conduit length [L]}
\nomenclature[Y]{$\mu$}{Dynamic viscosity [M L$^{-1}$ T$^{-1}$]}
\nomenclature[Z]{$n$}{Manning friction coefficient [T L$^{-1/3}$]}
\nomenclature[AA]{$P$}{Wetted perimeter [L]}
\nomenclature[AB]{$\Pc$}{Conduit perimeter [L]}
\nomenclature[AC]{$Q$}{Discharge in a conduit [L$^3$ T$^{-1}$]}
\nomenclature[AD]{$\Qt$}{Total nodal flow rate [L$^3$ T$^{-1}$]}
\nomenclature[AE]{$Q_f$}{Boundary flux per unit width [L$^2$ T$^{-1}$]}
\nomenclature[AF]{$Q_{\min}$}{Minimum flow discharge [L$^3$ T$^{-1}$]}
\nomenclature[AG]{$q$}{Areal recharge per unit length [L$^2$ T$^{-1}$]}
\nomenclature[AH]{$r_0$}{Uniform recharge rate [L T$^{-1}$]}
\nomenclature[AI]{$R$}{Hydraulic radius [L]}
\nomenclature[AJ]{$\Rc$}{Conduit radius [L]}
\nomenclature[AK]{$\R$}{Reynolds number [--]}
\nomenclature[AL]{RMSE}{Root mean square error [--]}
\nomenclature[AM]{$\rho$}{Water density [M L$^{-3}$]}
\nomenclature[AN]{$\Sb$}{Conduit bed slope [--]}
\nomenclature[AO]{$\Sf$}{Friction slope [--]}
\nomenclature[AP]{$S_i$}{Nodal free-surface area [L$^2$]}
\nomenclature[AQ]{$t$}{Time [T]}
\nomenclature[AR]{$v$}{Flow velocity [L T$^{-1}$]}
\nomenclature[AS]{$V_i$}{Nodal volume [L$^3$]}
\nomenclature[AT]{$W$}{Free-surface width [L]}
\nomenclature[AU]{$W_0$}{Preissmann slot width [L]}
\nomenclature[AV]{$x$}{Distance [L]}
\nomenclature[AW]{$y$}{Water depth in the conduit [L]}
\nomenclature[AX]{$\hat{y}$}{Normalized water depth [--]}
\nomenclature[AY]{$y_{\min}$}{Minimum water depth [L]}
\nomenclature[AZ]{$z$}{Conduit base elevation [L]}
\nomenclature[BA]{$\omega$}{Relaxation factor [--]}

\printnomenclature

\end{document}